\documentclass[12pt]{iopart}
\usepackage{graphicx}
\usepackage{dcolumn}
\usepackage{bm}
\usepackage[dvips]{color}

\def\la{\mathrel{\mathpalette\fun <}}
\def\ga{\mathrel{\mathpalette\fun >}}
\def\fun#1#2{\lower3.6pt\vbox{\baselineskip0pt\lineskip.9pt
  \ialign{$\mathsurround=0pt#1\hfil##\hfil$\crcr#2\crcr\sim\crcr}}}

 \definecolor{Black}{named}{Black}
 \definecolor{Blue}{named}{Blue}
 \definecolor{Red}{named}{Red}

\begin{document}

\title[Dark Matter Signatures in the Anisotropic Radio Sky]{Dark Matter Signatures in the Anisotropic Radio Sky}
\author{Le Zhang\dag, G\"unter Sigl\dag}
\address{\dag II. Institut f\"ur theoretische Physik, Universit\"at Hamburg,
Luruper Chaussee 149, D-22761 Hamburg, Germany}

\begin{abstract}
We calculate intensity and angular power spectrum of the cosmological background of
synchrotron emission from cold dark matter annihilations into electron positron pairs. We compare this background
with intensity and anisotropy of astrophysical and cosmological radio backgrounds, such
as from normal galaxies, radio-galaxies, galaxy cluster accretion shocks, the cosmic
microwave background and with Galactic foregrounds.
Under modest assumptions for the dark matter clustering we find that around 2 GHz
average intensity and
fluctuations of the radio background at sub-degree scales allows to probe dark matter
masses $\ga100\,$GeV and annihilation cross sections not far from the natural values
$\left\langle\sigma v\right\rangle\sim3\times10^{-26}\,{\rm cm}^3\,{\rm s}^{-1}$ required
to reproduce the correct relic density of thermal dark matter. The angular power spectrum
of the signal from dark matter annihilation tends to be flatter than that from astrophysical
radio backgrounds. Furthermore, radio source counts have comparable constraining power.
Such signatures are interesting especially for future radio detectors such as SKA.
\end{abstract}


\maketitle

\section{Introduction} 
Whereas it is known from cosmological observations that cold dark matter represents
a fraction $\Omega_m\simeq0.233$ of the total present energy density of our
Universe~\cite{Komatsu:2008hk,Dunkley:2008ie}, its nature is still elusive.
Dark matter can not only be detected directly in dedicated experiments searching for nuclear
recoils from the scattering of dark matter particles, or produced in particle accelerators such
as the LHC, but can also reveal its existence indirectly~\cite{Bertone:2004pz}: Although, apart from dilution
from cosmic expansion, the density of dark matter does not change significantly after
self-annihilations freeze out in the early Universe, residual self-annihilation can
give rise to significant fluxes of $\gamma-$rays, electrons, positrons, neutrinos, and even
some antimatter such as anti-protons and positrons, especially in regions with large dark matter
densities. The energies of the secondary particles can reach up to the dark matter particle mass
which can be of order a few hundred GeV. Secondary electrons and positrons can annihilate and
give rise to a 511 keV line emission, and they emit synchrotron radiation in the magnetic fields
of galaxies which can be detected in the radio band. Therefore, cosmic and $\gamma-$ray
detectors, neutrino telescopes, and even radio telescopes can be used for indirect dark matter
detection as well.

One of the most promising dark matter candidates are weakly interacting massive particles (WIMP)
such as they are predicted within supersymmetric extensions of the Standard Model. Such particles
have masses $m_X\ga100\,$GeV. If they are produced thermally, in order to reproduce the correct
average dark matter density, their annihilation cross sections have to be
$\left\langle\sigma v\right\rangle\sim3\times10^{-26}\,{\rm cm}^3/$s whereas large values are
possible in case of non-thermal production.

Traditionally, indirect dark matter detection has focused on high energy emission,
specifically signatures in $\gamma-$rays~\cite{Bergstrom:2001jj,Ullio:2002pj,Taylor:2002zd,Elsaesser:2004ap,Ando:2005xg,Ando:2006cr,Hooper:2007gi,Cuoco:2007sh,Ando:2006mt,Mack:2008wu}. High energy emission of neutrinos, the particle the most difficult to detect,
can be used to establish conservative constraints on the total annihilation cross section,
which are of the order $\sim10^{-23}\,{\rm cm}^3/$s~\cite{Yuksel:2007ac}. The fluxes of antiprotons
and positrons from galactic dark matter annihilations has also been extensively used to constrain dark matter properties~\cite{Bergstrom:2006tk,Bringmann:2006im,Delahaye:2007fr}. Furthermore, the synchrotron radiation
emitted by dark matter annihilation products close to the Galactic centre has also been studied.
It has been found that if the dark matter profile close to the central black hole is a spike
formed by adiabatic accretion, typical dark matter annihilation cross sections within supersymmetric
scenarios can lead to intensities comparable to the radio emission observed from the Galactic
centre~\cite{Bertone:2001jv,Bertone:2002je}. It has further been shown that measurements of the
radio flux away from the Galactic
centre by the WMAP experiment strongly constrain the annihilation cross sections to values
$\la10^{-25}\,{\rm cm}^3/$s for $m_X\simeq100\,$GeV~\cite{Hooper:2008zg,Grajek:2008jb}.
A multi-wavelength analysis of dark matter annihilations from the Galactic centre has recently been performed in
Ref.~\cite{Regis:2008ij}.

In the present paper we evaluate the diffuse synchrotron emission from the electrons
and positrons produced by dark matter annihilation in the cosmological distribution of
dark matter halos. We compute both its overall intensity and its angular power spectrum
as well as the distribution of visible dark matter annihilation sources
as a function of apparent luminosity. We will
find that comparing the resulting signals with other backgrounds and foregrounds under conservative
assumptions allows to test annihilation cross sections close to the natural scale
$\left\langle\sigma v\right\rangle\sim3\times10^{-26}\,{\rm cm}^3/$s.

In Sect.~2 we provide the general setup of our calculations. In Sect.~3 and~4,
we apply it to astrophysical backgrounds and the dark matter induced signal, respectively. In
Sect.~5 we compare the overall diffuse signal and its anisotropy to other foregrounds and discuss
the resulting dark matter constraints, and in Sect.~6 we conclude. Finally, an appendix presents
technical details of the calculations.
We will use natural units in which $c=1$ throughout.

\section{Setup}
We consider a distribution of sources which emit a radio luminosity per
frequency interval $L(\nu,{\cal P},z)$ which depends on a parameter ${\cal P}$,
on frequency $\nu$ and on redshift $z$. The energy flux per frequency
interval and solid angle is then given by
\begin{equation}\label{eq:I0}
  J(\nu)=\int dz\frac{d^2V}{dzd\Omega}\int d{\cal P}\frac{dn}{d{\cal P}}({\cal P},z)
  \frac{(1+z)L[\nu_z,{\cal P},z]}{4\pi d_L(z)^2}\,,
\end{equation}
where for abbreviation we write $\nu_z\equiv(1+z)\nu$,
$(dn/d{\cal P})({\cal P},z)$ is the co-moving volume density of objects per unit interval
in the parameter ${\cal P}$, $d_L(z)$ is the luminosity distance, the factor $1+z$
comes from redshifting the frequency interval $d\nu$, and the co-moving volume
per solid angle and redshift interval is
\begin{equation}\label{eq:dV}
  \frac{d^2V}{dzd\Omega}=\frac{d_L(z)^2}{(1+z)^2H(z)}=\frac{r(z)^2}{H(z)}\,.
\end{equation}
Here, for a flat cosmological geometry, the Hubble rate is
\begin{equation}\label{eq:cosmo}
  H(z)= H_0\left[\Omega_{nr}(1+z)^3+\Omega_{\Lambda}\right]^{1/2}\,,
\end{equation}
$r(z)=\int_{t(z)}^{t(0)}(1+z)dt=\int dz^\prime/H(z^\prime)$ is the co-moving distance
and $t(z)=\int_0^z dz^\prime/[(1+z^\prime)H(z^\prime)]$ is cosmic time as function of
redshift. Throughout this paper we will assume a flat, $\Lambda$CDM Universe
with the total non-relativistic matter density $\Omega_{nr}=0.279$ and the dark energy density
$\Omega_{\Lambda}=1-\Omega_{nr}\simeq0.701$ (all other
contributions to the energy density are negligible) in units of the critical energy density
$\rho_c=3H_0^2/(8\pi G_{\rm N})$, where $G_{\rm N}$ is Newtons constant and
$H_0=H(0)=100~h~{\rm km}~{\rm s}^{-1}~{\rm Mpc}^{-1}$ with $h=0.701$~\cite{Komatsu:2008hk,Dunkley:2008ie}.

In order to calculate the anisotropies we introduce an emissivity of
squared power per frequency interval, ${\cal L}^2(\nu,k,z)$,
\begin{equation}\label{eq:L2}
  {\cal L}^2(\nu,k,z)={\cal L}_1^2(\nu,k,z)+{\cal L}_2^2(\nu,k,z)\,,
\end{equation}
which, similar to the approach in Ref.~\cite{Ando:2006cr}, we split
into the two parts ${\cal L}_1^2(\nu,k,z)$ and
${\cal L}_2^2(\nu,k,z)$ and which also depends on the co-moving wavenumber $k$.
The first part is essentially Poisson noise
and corresponds to the sum over squared luminosities,
\begin{equation}\label{eq:L21}
  {\cal L}_1^2(\nu,k,z)=\int d{\cal P}\frac{dn}{d{\cal P}}({\cal P},z)
  \left[L(\nu_z,{\cal P},z)|u(k,{\cal P})|\right]^2\,,
\end{equation}
where in the following we define ${\cal F}_f(k)\equiv\int d^3{\bf r}e^{i{\bf k}\cdot{\bf r}}f({\bf r})$
as the spatial Fourier transform of any function $f({\bf r})$ and where $u(k,{\cal P})={\cal F}_u(k,{\cal P})$
is the Fourier transform of the spatial emission density $u({\bf r},{\cal P})$
of an individual source, normalized to unity, $\int d^3{\bf r}u({\bf r},{\cal P})=1$.
The second contribution to Eq.~(\ref{eq:L2}) is determined by the correlation
between sources,
\begin{equation}\label{eq:L22}
  {\cal L}_2^2(\nu,k,z)=P_{\rm lin}(k,z)
  \left[\int d{\cal P}\frac{dn}{d{\cal P}}({\cal P},z)
  L(\nu_z,{\cal P},z)b({\cal P},z)|u(k,{\cal P})|\right]^2\,,
\end{equation}
where $P_{\rm lin}(k,z)=\int d^3{\bf r}e^{i{\bf k}\cdot{\bf r}}(\delta\rho/\rho)({\bf r},z)$
is the linear power spectrum of the density fluctuations $(\delta\rho/\rho)({\bf r},z)$
and we have also introduced a bias factor $b({\cal P},z)$ of the sources
with respect to the density field.

The angular power spectrum $C_l$ is given by
\begin{equation}\label{eq:Cl}
  C_l=\left\langle|a_{lm}|^2\right\rangle\,,
\end{equation}
where
\begin{equation}\label{eq:alm}
  a_{lm}=\int d\Omega\left[J(\nu,\Omega)-\left\langle   
  J(\nu)\right\rangle\right]Y^*_{lm}(\Omega)
\end{equation}
in terms of the spherical harmonic functions $Y_{lm}(\Omega)$ and the intensity
$J(\nu,\Omega)$ measured along direction $\Omega$. For
a statistically isotropic sky this results in
\begin{equation}\label{eq:Cl2}
  C_l=\int dz\frac{d^2V}{dzd\Omega}
  \frac{(1+z)^2{\cal L}^2\left(\nu_z,\frac{l}{r(z)},z\right)}
  {[4\pi d_L(z)^2]^2}\,.
\end{equation}
Using $d_L(z)=(1+z)r(z)$ and inserting Eq.~(\ref{eq:dV}) in Eqs.~(\ref{eq:I0}) and~(\ref{eq:Cl2})
finally gives
\begin{equation}\label{eq:I02}
  J(\nu)=\frac{1}{4\pi}\int\frac{dz}{(1+z)H(z)}\int d{\cal P}
  \frac{dn}{d{\cal P}}({\cal P},z)L(\nu_z,{\cal P},z)\,,
\end{equation}
and
\begin{equation}\label{eq:Cl22}
  C_l=\frac{1}{(4\pi)^2}\int dz
  \frac{{\cal L}^2\left(\nu_z,\frac{l}{r(z)},z\right)}
  {d_L(z)^2H(z)}\,.
\end{equation}
Formally, for point-like sources, the integral over redshift in Eq.~(\ref{eq:Cl22}) is
divergent at $z\to0$. In practice this is regularized by the fact that the nearest source has
some minimal distance and that one can subtract the most luminous point sources
which are also the nearest sources. In addition, the integral is regularized by the
spatial extent of the sources, represented by the factor $|u(k,{\cal P})|^2$ in Eqs.~(\ref{eq:L21})
and~(\ref{eq:L22}). The role of these effects in practical calculations will be discussed in
Sect.~5.2.

\section{Astrophysical Sources}
For astrophysical sources, ${\cal P}$ can be identified with the
radio luminosity $L_{\nu_0}$ at some fixed frequency $\nu_0$.
Eq.~(\ref{eq:I02}) then simplifies to
\begin{equation}\label{eq:I0_astro}
  J(\nu)=\frac{1}{4\pi}\int\frac{dz}{(1+z)H(z)}\int^{L_{\rm cut}(z)} dL_{\nu_0}
  L_{\nu_0}\frac{dn}{dL_{\nu_0}}(L_{\nu_0},z)
  \frac{L(\nu_z)}{L_{\nu_0}}\,,
\end{equation}
where $L_{\rm cut}(z)=4\pi d_L(z)^2S_{\rm cut}/(1+z)$ is the intrinsic luminosity corresponding to
the apparent point source flux $S_{\rm cut}$ above which we consider the source to
be resolvable and thus subtractable from the diffuse background.
For the multipoles Eqs.~(\ref{eq:L21}) and~(\ref{eq:L22}) can then be written as
\begin{equation}\label{eq:L21_astro}
  \fl{\cal L}_1^2(\nu,k,z)=\int^{L_{\rm cut}(z)}
  dL_{\nu_0}L_{\nu_0}^2\frac{dn}{dL_{\nu_0}}(L_{\nu_0},z)
  \left[\frac{L(\nu_z)}{L_{\nu_0}}u(k,z)\right]^2
\end{equation}
and
\begin{equation}\label{eq:L22_astro}
  \fl{\cal L}_2^2(\nu,k,z)=P_{\rm lin}(k,z)
  \left[\int^{L_{\rm cut}(z)} dL_{\nu_0}L_{\nu_0}\frac{dn}{dL_{\nu_0}}(L_{\nu_0},z)
  \frac{L(\nu_z)}{L_{\nu_0}}u(k,z)b(L_{\nu_0},z)\right]^2\,,
\end{equation}
respectively. For the luminosity functions $dn/dL_{\nu_0}$ of normal and radio galaxies
we will use the expressions given in Ref.~\cite{Protheroe:1996si}.

\section{Dark Matter Annihilation}
For annihilation of dark matter with mass $m_X$ and phase space averaged annihilation
cross section times velocity $\left\langle\sigma v\right\rangle$, ${\cal P}$ can be identified
with the mass $M$ of dark matter halos. We then follow the approach of
Ref.~\cite{Bertone:2001jv} and write
\begin{equation}\label{eq:L_dm}
  L(\nu,M)=\frac{\left\langle\sigma v\right\rangle}{2m_X^2}\,{\cal E}(\nu,M)\,,
\end{equation}
where we define ${\cal E}(\nu,M)$ as a quantity which does not depend on
annihilation cross section or mass of the dark matter particles,
\begin{equation}\label{eq:E_ann}
  \fl{\cal E}(\nu,M)=\frac{\sqrt{3}e^3}{m_e}\int d^3{\bf r}\rho^2_h({\bf r})
  B({\bf r})\int_{m_e}^{m_X}dE\frac{Y_e(>E)}{P_{\rm syn}(E)+P_{\rm IC}(E)}
  F\left[\frac{\nu}{\nu_c(E)}\right]\,.
\end{equation}
In Eq.~(\ref{eq:E_ann}), $e$ and $m_e$ are the electron charge and mass,
respectively, $\rho_h({\bf r})$ is the dark matter halo density profile,
$B({\bf r})$ is the local magnetic field strength, , $Y_e(>E)$
is the multiplicity per annihilation of electrons and positrons with energies larger
than $E$, $P_{\rm syn}(E)=2e^4B^2E^2/(3m_e^4)=(16e^4\pi/3)u_B E^2/m_e^4$
is the total synchrotron emission power of one electron of energy $E$ in a
magnetic field of strength $B$, corresponding to an energy density $u_B=B^2/(8\pi)$,
and $P_{\rm IC}(E)=(16e^4\pi/3)u_\gamma E^2/m_e^4$ is the energy loss rate at
energy $E$ due to inverse Compton scattering on a low energy photon field of
energy density $u_\gamma$. Furthermore, we use the function
\begin{equation}\label{eq:F_synch}
  F(x)=x\int_x^{\infty} K_{5/3}(y) dy\,,
\end{equation}
in Eq.~(\ref{eq:E_ann}), with the critical frequeny
\begin{equation}\label{eq:nu_crit}
  \nu_c(E)=\frac{3}{4\pi}\frac{eB}{m_e}\left(\frac{E}{m_e}\right)^2\,.
\end{equation}

In the following we use the approximation~\cite{Rybicki}
\begin{equation}\label{eq:nu_crit_approx}
  F(x)\simeq\delta[x-0.29]
\end{equation}
such that Eq.~(\ref{eq:E_ann}) can be simplified to
\begin{equation}\label{eq:E_ann2}
  {\cal E}({\nu},M)\simeq\frac{9}{8}\left(\frac{m_e^3}{0.29\pi}\right)^{1/2}
  \frac{Y_e[>E_c(\nu)]}{\nu^{1/2}}\,I(M)
\end{equation}
where 
\begin{equation}\label{eq:I_m}
  I(M)=\int d^3{\bf r}\frac{\rho^2_m({\bf r})}{(eB)^{1/2}({\bf r})}
  \frac{1}{1+u_\gamma({\bf r})/u_B({\bf r})}\,,
\end{equation}
and the critical energy $E_c(\nu)$ is the inversion of Eq.~(\ref{eq:nu_crit}),
\begin{equation}\label{eq:E_crit}
  E_c(\nu)=\left(\frac{4 \pi}{3\cdot0.29}\frac{m_e^3}{e}\frac{\nu}{B}\right)^{1/2}=
  5.9\left(\frac{\nu}{1\,{\rm GHz}}\right)^{1/2}\left(\frac{B}{6\,\mu{\rm G}}\right)^{-1/2}
  \,{\rm GeV}\,.
\end{equation}
In Eq.~(\ref{eq:E_ann2}) we neglect the magnetic field dependence of $Y_e[>E_c(\nu)]$.
For $m_X\ga100\,$GeV, $B\ga$ a few micro-Gauss and $\nu\sim1\,$GHz,
the parameters we are interested in, this is a good approximation because the critical energy
$E_c(\nu_z)\la m_X/10$. Typical values for these parameters are $Y_e\simeq10$~\cite{Bertone:2001jv}.
This corresponds to a fraction $f_e\simeq0.3$ of the total annihilation energy
going into pairs. The energy fraction going into pairs of energy above $E$ can be expressed
in terms of $Y_e(>E)$ as
\begin{equation}\label{eq:f_e}
  f_e(E)=\frac{-1}{2m_X}\int_{E}^{m_X}dE^\prime E^\prime\frac{dY_e}{dE}(E^\prime)\leq1\,.
\end{equation}

With the above expressions we can rewrite Eq.~(\ref{eq:I02}) as
\begin{equation}\label{eq:I0_dm}
  \fl J(\nu)=\frac{\left\langle\sigma v\right\rangle}{2m_X^2}\frac{9}{32\pi}
  \left(\frac{m_e^3}{0.29\pi\nu}\right)^{1/2}\int\frac{dz}{(1+z)^{3/2}H(z)}\int dM
  \frac{dn}{dM}(M,z)Y_e[>E_c(\nu_z)]I(M)\,.
\end{equation}
Furthermore, we can redefine ${\cal L}_1^2$ and ${\cal L}_2^2$ from
Eqs.~(\ref{eq:L21}) and~(\ref{eq:L22}) by extracting constant factors
and write
\begin{equation}\label{eq:L21_dm}
  \fl {\cal L}_1^2(\nu,k,z)=\int dM\frac{dn}{dM}(M,z)
  \left(Y_e[>E_c(\nu_z)]I(M)|u(k,M)|\right)^2
\end{equation}
and
\begin{equation}\label{eq:L22_dm}
  \fl {\cal L}_2^2(\nu,k,z)=P_{\rm lin}(k,z)\left(\int dM\frac{dn}{dM}(M,z)
  Y_e[>E_c(\nu_z)]I(M)b(M,z)|u(k,M)|\right)^2\,,
\end{equation}
where $u(k,M)$ relates to the halo profile, $u(k,M)=\int d^3{\bf r}e^{i{\bf k}\cdot{\rm r}}
\rho^2_h({\bf r})(eB)^{-1/2}({\bf r})/I(M)$. With these quantities we can now write
\begin{equation}\label{eq:Cl_dm}
  C_l=\frac{81m_e^3}{1024\cdot0.29\pi^3\nu}\left(\frac{\sigma v}{m_X^2}\right)^2
  \int dz\frac{{\cal L}_1^2\left(\nu,\frac{l}{r(z)},z\right)+
  {\cal L}_2^2\left(\nu,\frac{l}{r(z)},z\right)}{(1+z)d_L(z)^2H(z)}\,.
\end{equation}
Details about the quantities that enter these expressions are given in Appendix~A.

Eq.~(\ref{eq:Cl_dm}) can also be obtained as follows: Limber's equation relates the two-dimensional
angular power spectrum $P_2(l)$ to the three-dimensional power spectrum $P_3(k)$ in the flat sky approximation~\cite{kaiser:1992apj}: Given a three-dimensional statistically random field
$f({\bf r})=f(\Omega,r)$, one considers the observation at ${\bf r}=0$ of the projection 
\begin{equation}\label{eq:limber1}
  P(\Omega)=\int dr ~w(r) ~f(\Omega,r)
\end{equation}
with some given radial weight function $w(r)$, where $r$ is the co-moving distance. If the
field $f$ fluctuates on scales much smaller than the characteristic scale over which $w(r)$
varies, then we have 
\begin{equation}\label{eq:limber2}
  C_l\simeq\int dr \frac{w^2(r)}{r^2} P_f(l/r,z(r))
\end{equation}
where $P_f(l/r,z)$ is the power spectrum of $\left\langle f(\Omega_1,r)f(\Omega_2,r)\right\rangle$
at the co-moving wavenumber $k=l/r$.

Neglecting the variation of the magnetic field $B$ within the halo regions contributing most
to the annihilations, the radio intensity Eq.~(\ref{eq:I0_dm}) along a given direction $\Omega$
can be written as
 \begin{equation}\label{eq:I-syn-dm1}
  \fl J(\nu,\Omega)=\frac{\left\langle\sigma v\right\rangle}{m_X^2}\frac{9\rho_m}{64\pi\nu^{1/2}}
  \left(\frac{m_e^3}{0.29\pi eB}\right)^{1/2}\int dz\frac{(1+z)^{3/2}}{H(z)}Y_e[>E_c(\nu_z)]   
  \frac{\left[1+\delta(z,\Omega)\right]^2}{1+u_\gamma/u_B}\,,
\end{equation}
where $\rho_m=\Omega_m\rho_c$ is the average dark matter density at zero redshift,
and $\delta=\delta\rho/\rho$ is the relative overdensity. Because the dominant contribution comes from
the dark matter halos, where $\delta\gg1$, we can approximate $(1+\delta)^2\simeq\delta^2$.
Assuming a constant $B$ and a constant optical photon field of density
$u_{\rm op}\simeq5\,{\rm eV}\,{\rm cm}^{-3}$, we can write the factor 
$(1+u_\gamma/u_B)^{-1}=\left[1+u_{\rm op}/u_B+u_0(1+z)^4/u_B\right]^{-1}$,
where $u_0$ is the CMB energy density at $z=0$. This factor effectively cuts off the redshift
integration at $z\simeq2$. Since $E_c(\nu_z)$ varies little over this redshift range,
we can then further simplify Eq.~(\ref{eq:I-syn-dm1}) to
\begin{equation}\label{eq:I-syn-dm2}
  \fl J(\nu,\Omega)\simeq \frac{Y_e[>E_c(\nu)]\left\langle\sigma v\right\rangle}{m_X^2}
  \frac{9\rho^2_m}{64\pi\sqrt{eB\nu}}
  \left(\frac{m_e^3}{0.29\pi}\right)^{1/2}\int dz
  \frac{(1+z)^{3/2}\delta^2(z,\Omega)}{H(z)\left[1+\frac{u_{\rm op}}{u_B}+\frac{u_0}{u_B}(1+z)^4\right]}\,.
\end{equation}
Comparing this with Eq.~(\ref{eq:limber1}), we can use
\begin{equation}\label{eq:I-syn-dm2-f}
f = \delta^2-\left\langle\delta^2\right\rangle\,,
\end{equation}
for the random field and the weight function is
\begin{equation}\label{eq:I-syn-dm2-w}
  w(z)=\frac{Y_e[>E_c(\nu)]\left\langle\sigma v\right\rangle}{m_X^2}
  \frac{9\rho^2_m}{64\pi\sqrt{eB\nu}}
  \left(\frac{m_e^3}{0.29\pi}\right)^{1/2}\frac{(1+z)^{3/2}}{1+\frac{u_{\rm op}}{u_B}+\frac{u_0}{u_B}(1+z)^4}\,.
\end{equation}
The power spectrum $P_f(k,z)$ appearing in Eq.~(\ref{eq:limber2}) is then the Fourier transform of the
two-point correlation function of $f$ in real space. Following Ref.~\cite{Ando:2005xg}, $P_{\delta^2}(k,z)$
can be written as the sum of a one-halo and a two-halo term,
$P_{\delta^2}(k,z)=P_{\delta^2}^{1h}(k,z)+P_{\delta^2}^{2h}(k,z)$,
with
\begin{equation}\label{eq:I-syn-dm2-pk1h}
  \fl P_{\delta^2}^{1h}(k,z)=\int_{M_{min}}^{M_{\rm cut}(z)}dM\frac{dn}{dM}
  \left[{\cal F}_{\delta^2}(k,M,z)\right]^2=
  \int_{M_{min}}^{M_{\rm cut}(z)} dM \frac{dn}{dM}\left(\frac{A_b{\cal F}_{\rho^2_h}(k,M,z)}
  {\rho_m^2(1+z)^6}\right)^2
\end{equation}
\begin{eqnarray}\label{eq:I-syn-dm2-pk2h}
  \fl P_{\delta^2}^{2h}(k,z)&=&P_{\rm lin}(k)\left[\int_{M_{min}}^{M_{\rm cut}(z)} dM \frac{dn}{dM} b(M)
  {\cal F}_{\delta^2}(k,M,z)\right]^2\nonumber\\
  \fl &=&P_{\rm lin}(k)\left[\int_{M_{min}}^{M_{\rm cut}(z)}
  dM \frac{dn}{dM} b(M)\left(\frac{A_b{\cal F}_{\rho^2_h}(k,M,z)}{\rho_m^2(1+z)^6}\right)
  \right]^2\,,
\end{eqnarray}
where $M_{\rm min}$ is the minimal halo mass and $M_{\rm cut}(z)$ is the halo mass 
corresponding to the apparent point source flux $S_{\rm cut}$ above which we consider the source to
be resolvable and thus subtractable from the diffuse background. Furthermore,
$A_b$ is a boost factor which accounts for possible
substructure in the halos. The average of the clumping factor appearing in
Eq.~(\ref{eq:I-syn-dm2}) is given by
\begin{equation}\label{eq:I-syn-dm<>}
  \left\langle\delta^2(z)\right\rangle= \frac{A_b}{\rho^2_m(1+z)^6}
  \int_{M_{\rm min}}^{M_{\rm cut}(z)} dM \frac{dn}{dM}\times \int dV_h\,\rho^2_h({\rm r},M,z)\,,
\end{equation}
where $dV_h$ is the halo volume element.

A generic form for the halo mass function $dn/dM$ appearing in the equations above was first proposed by
Press \& Schechter (PS)~\cite{Press:1973iz}; a modified version of this form is given by Sheth and 
Tormen~\cite{Sheth:1999mn} (ST). When comparing the results obtained from these two forms, we find differences by factors less than 2. Thus, we adopt the PS formula throughout our paper.

Current knowledge of the dark matter density distribution mostly comes from
N-body simulations, and the universal dark matter profile firstly proposed is the Navarro-Frenk-White
(NFW) model~\cite{Navarro:1995iw}. Combining Eq.~(\ref{eq:rho_NFW}) with Eq.~(\ref{eq:delta_ch}) in Appendix~A,
in this model the dark matter profile within each halo can be written as
\begin{equation}\label{eq:NFW}
  \rho_h(r) = \frac{\Delta_c(z)}{3}\frac{c^3}{\ln(1+c)-c/(1+c)}\,\frac{\rho_m(z)}{r/r_s(1+r/r_s)^2}\,,
\end{equation}
where $r_s$ is a characteristic radius, and the concentration parameter $c$ is defined as the
ratio of $r_s$ and the virial radius $r_v$, $c\equiv {r_v}/{r_s}$, see also Appendix A for more
details. Note that $r_s$ is not a free parameter,
but depends on $M$ and $c$ because $r_v$ is related to $M$ via $M=4\pi\Delta_c(z)\rho_m(z)/3$,
where $\Delta_c(z)$ in an Einstein-de Sitter Universe is about $18\pi^2$. With the above definition, $c$ and $M$ completely determine the dark matter distribution of a given halo. The minimal halo mass is still
rather uncertain. The value $M_{\rm min}=10^{-6}\,M_\odot$~\cite{Diemand:2005vz} is close to the free-streaming mass~\cite{hofmann01,chen01,green05}, below which there are no fluctuations in the dark
matter density to form a halo. Note that the magnetic field may be much smaller than micro Gauss
scales in such small halos. In contrast, the value $M_{\rm min}=10^{6}\,M_\odot$ roughly corresponds
to the minimal mass of dwarf galaxies which are known to contain micro Gauss scale magnetic 
fields~\cite{Kronberg:1993vk}. We, therefore,
choose $M_{\rm min}=10^{6}\,M_\odot$ as fiducial value in the following, noting that the dark matter
signal would increase by only a factor about two for $M_{\rm min}=10^{-6}\,M_\odot$. We will furthermore use $B=10\mu$G as fiducial value for the magnetic field. This is a realistic
value given that most annihilations occur in the densest regions
where also magnetic fields are somewhat larger than typical average galactic fields.

The clumping factor is very sensitive to the concentration parameter, namely $\propto c^3$.
N-body simulations indicate that the concentration has a log-normal distribution~\cite{Bullock:1999he}
with a median value of
\begin{equation}\label{eq:con}
  c(M,z)=4\,\frac{1+z_c}{1+z}\,,
\end{equation}  
where the collapse redshift $z_c$ is implicitly given by the relation $M_*(z_c)=0.01M$, where $M_*(z)$ is the mass scale at which $\sigma(M_*,z)=\delta_c$. How the concentration parameter depends on halo mass and redshift is
still an open question. One can extrapolate Eq.~(\ref{eq:con}) to minimal halo masses $M\sim10^{-6}\,M_\odot$.
When comparing the parameterization Eq.~(\ref{eq:con}) with high resolution simulations~\cite{Diemand:2005vz} we find that it gives realistic values for the minimum halo mass.
It is a conservative estimate because at $z\simeq0$ it gives values $c\sim 70$ for the minimum halo mass
which is significantly smaller
than other parameterizations~\cite{Seljak:2000gq,Cooray:2002dia}.

Recent studies show that dark matter halos exhibit considerable 
substructure~\cite{tormen98,klypin99,moore99,ghigna00,springel01,zentner03,lucia04,helmi02,gao04,shaw06,Diemand:2008in}.
The total mass of these substructures only account for about $10\%$ of the host halo, but they can
give an extra boost factor $A_b\sim10$ for dark matter annihilation. Some studies show that if one takes
into account substructure and assumes a cuspy center slope~\cite{Bi:2005im,Yuan:2006ju,Lavalle:1900wn}, the theoretical prediction can well explain the excess of high energy positrons and the diffuse $\gamma-$
ray background observed by the Heat~\cite{Barwick:1997ig,Coutu:2001jy} and 
EGRET~\cite{Strong:2004ry,Sreekumar:1997un} experiments, respectively. The subhalos follow a certain mass and redshift distribution which is
still unknown. Therefore, to be conservative we assume the NFW halo model and simply parametrize
any possible boost factor with the parameter $A_b\sim10$. The substructures occur on
small scales and do not influence the power spectrum in the range we are interested, $l\la10^4$.

\section{Results}
\subsection{Diffuse Radio Emission}
The cosmic microwave background (CMB) dominates the radio sky at frequencies above $\simeq1\,$GHz,
whereas astrophysical sources such as normal galaxies and radio galaxies dominate at lower frequencies
down to kHz frequencies~\cite{Protheroe:1996si}. Recently it was argued that synchrotron emission
of strong intergalactic shocks can also significantly contribute to the diffuse extragalactic radio
below 500 MHz~\cite{Waxman:2000pf,Keshet:2004dr}.

Using the formuli developed in Sect.~4, we now evaluate the contribution of synchrotron emission
from pairs produced by dark matter annihilation in the magnetic fields of dark matter halos.
We consider neutralinos as dark matter candidate, and for the following figures we assume a neutralino
mass of $100$ GeV and a total annihilation cross section of
$\left\langle\sigma v\right\rangle=3\times10^{-26}\,{\rm cm}^3$/s, fixed
for reproducing the correct relic density for thermal relics. We also assume that the average
total number of electrons and positrons per annihilation is $Y_e\simeq10$, and that the halo substructure
implies a boost factor $A_b\simeq10$. We compare the resulting dark matter
signal with astrophysical contributions to the diffuse background that can be computed from
the expressions in Sect.~3.

For astrophysical sources the diffuse radio background is likely dominated by normal galaxies and radio
galaxies. To estimate the contributions from these sources, we follow Ref.~\cite{Protheroe:1996si},
which use the observed correlation between the radio and infra-red flux of galaxies. This approach
assumes that the radio emission is related to the star formation and is sensitive to the redshift
evolution of the sources, but can explain the observed radio background quite well.

\begin{figure}
\includegraphics[width=0.6\textwidth]{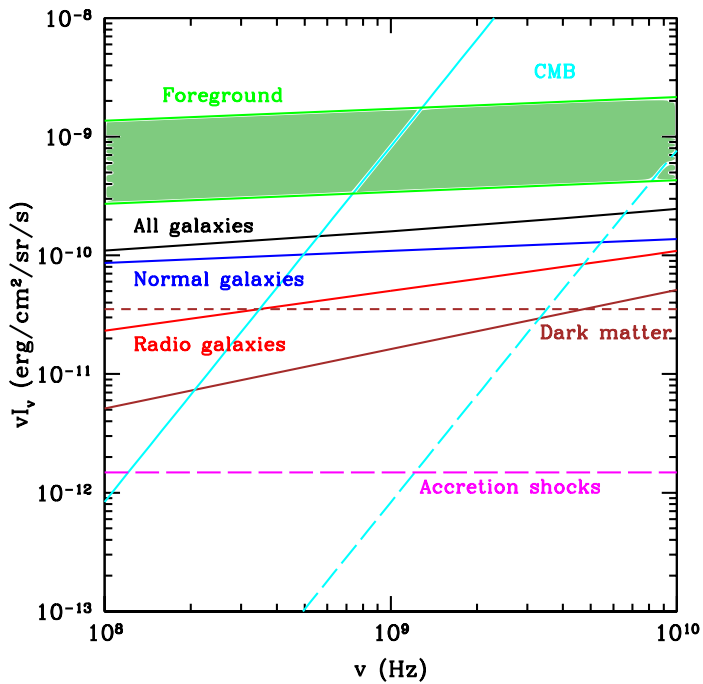}
\caption{The average diffuse background flux intensity with no point-source removal. Contributions from normal galaxies (blue curve), radio galaxies (red curve), from radio and normal galaxies combined (black curve),
and from a scenario for radio emission from galaxy cluster shocks (magenta curve)~\cite{Keshet:2004dr}
(see text for the normalization) are compared to our
fiducial dark matter annihilation scenario with $m_X=100\,$GeV,
$\left\langle\sigma v\right\rangle\sim3\times10^{-26}\,{\rm cm}^3/$s, $A_b=10$,
$B=10\,\mu$G, $M_{\rm min}= 10^{6}\,M_{\odot}$ (brown curves). Here, the solid brown curve
is for $Y_e=10$, while the dashed brown curve is for $Y_e(E)\simeq m_X/E$. Also shown is the
CMB background (cyan solid curve) as well as its subtractable part, determined by uncertainties of the absolute
CMB temperature (dotted cyan curve). The Galactic foreground at Galactic latitude
$b>20^\circ$ is shown as the green band within uncertainties.}
\label{fig:radio_bg}
\end{figure}

Following the above assumptions, in Fig.~\ref{fig:radio_bg}, we show the different contributions to
the average diffuse radio intensity. For astrophysics sources, normal galaxies contribute more than
radio galaxies. This is because although the individual radio galaxy is brighter than a normal galaxy
on average, this is overcompensated by the larger number of normal galaxies. Also shown
in Fig.~\ref{fig:radio_bg} is a possible contribution from intergalactic shocks~\cite{Waxman:2000pf,Keshet:2004dr}
normalized such that its angular power spectrum is comparable to the one of the Galactic foreground,
see Sect.5.2.

Of course, the CMB absolutely dominates the radio sky in the wide range from $\nu\simeq1\,$GHz to a few hundred
GHz~\cite{Smoot:1998jk}, and above these frequencies Galactic foregrounds such as dust emission dominates. Since the CMB is a black body radiator its contribution to the solid angle averaged radio flux can be subtracted
up to the uncertainty of its average absolute temperature. Currently the CMB temperature is measured
to $2.725 \pm 0.001$ K~\cite{Mather:1998gm}.
We convert this temperature uncertainty into an intensity of CMB confusion noise. Fig.~\ref{fig:radio_bg}
shows that this confusion noise dominates other astrophysical backgrounds and the diffuse signal 
of our fiducial dark matter scenario at $\nu\ga4\,$GHz. At lower frequencies the dark matter
signal $\nu J(\nu)$ tends to decrease as $\sqrt\nu$ for $Y_e\simeq\,$const., see Eq.~(\ref{eq:I-syn-dm2}),
whereas the background from normal galaxies tends to be flat, see Fig.~\ref{fig:radio_bg}.
There is thus an optimal window at frequencies $\nu\sim1\,$GHz where dark matter annihilation signatures
can be detected and where self-absorption is negligible. Constraints on dark matter
parameters can, therefore, only be established for annihilation cross sections about a factor ten
higher than the fiducial cross section required for thermal dark matter.

In addition, there are three diffuse foregrounds from our Galaxy in the frequency range we are interested:
The first is synchrotron radiation emitted by high energy electrons gyrating in the Galactic
magnetic field, the second is free-free emission from the thermal bremsstrahlung from hot ($\geq 10^4$K)
electrons produced in the interstellar gas by the Galactic UV radiation field, and the third foreground
is dust emission which arises from the thermal re-radiation of absorbed stellar light.
Fig.~\ref{fig:radio_bg} shows that these foregrounds tend to dominate the astrophysical
backgrounds and the dark matter signal in the fiducial scenario.

Can we test the properties of dark matter more powerfully? The absolute CMB temperature is difficult
to measure more precisely than to the current permille
level, because of inevitable systematic errors. Small-scale temperature fluctuations
$\Delta T/T\sim 10^{-5}$  have been seen by the COBE and WMAP satellites because temperature
differences can be measured more precisely since systematic errors cancel in measurements of temperature differences.
Furthermore, if the Galactic foregrounds have a smooth directional dependence, they may pose less
of a contamination when considering the anisotropy of the radio sky. We,
therefore, consider in the following the angular power spectra of the radio sky in order to see if
it can provide further tests of dark matter properties.

\subsection{Anisotropy}
Whereas the diffuse average radio flux provides only one number at a given frequency to compare
with other astrophysical and cosmological backgrounds, potentially much more information is contained
in the angular power spectrum. For example, the power spectrum as a function of angular scale tends
to be different for dark matter annihilation and astrophysical sources because the contribution of
the latter to the diffuse radio flux is dominated by fewer bright sources. Our goal in this section is
whether this can provide dark matter signatures or constraints on mass and annihilation cross section.

Before calculating the angular power spectra, we discuss their qualitative behaviors. The angular power spectrum $C_l=C_l^{1h}+C_l^{2h}$ can be divided into one-halo ($C_l^{1h}$) and two-halo ($C_l^{2h}$) terms, corresponding
to the two contributions Eqs.~(\ref{eq:L21}) and (\ref{eq:L22}) to Eq.~(\ref{eq:L2}), and thus to
Eq.~(\ref{eq:Cl22}). The two-halo term arises from the correlation between distinct halos which is described by
the linear power spectrum. The one-halo term represents correlation within the same halo. Both one-halo
and two-halo term are proportional to $|u(k,{\cal P})|^2$, the square of the Fourier transform of
the spatial emission profile. At large angular scales, $|u(k,{\cal P})|\sim1$, such that $C_l^{1h}$ is
essentially independent of $l$. The one-halo term is thus sometimes called Poisson noise. At scales comparable to
the size of the source, $|u(k,{\cal P})|^2$ starts to become suppressed. Therefore, both one-halo and
two-halo terms are expected to be suppressed for multipoles $l$ larger than the typical distance to
the source divided by the linear source size. The two-halo term is furthermore proportional to the
linear power spectrum which is also suppressed for co-moving wavenumbers $k\ga0.03\,{\rm Mpc}^{-1}$. Therefore, the
ratio of the two-halo term to the one-halo term is suppressed for $l\ga0.03\,{\rm Mpc}^{-1}r_{\rm H}\simeq100$,
where $r_{\rm H}\simeq3000\,$Mpc is the Hubble scale. The one-halo term eventually dominates at
very small angular scales.

In Eq.~(\ref{eq:Cl22}), for point-like sources, formally the one-halo term $C_l^{1h}$ would diverge for
$z_{\rm min}\to0$, whereas the two-halo term Eq.~(\ref{eq:L22}) is regularized by the linear power
spectrum $P_{\rm lin}(k,z)$, which is suppressed at large $k=l/r(z)$. This is because the flux of
nearby sources of a given luminosity diverges.
We can ignore such sources because they can be identified as individual bright sources
and be removed from the background flux in actual observations. We can remove sources
with intrinsic luminosity $L_{\rm cut}(z)\geq4\pi d_L(z)^2S_{\rm cut}/(1+z)$, corresponding to
the point-source sensitivity $S_{\rm cut}$ of the
telescope. Alternatively, one can regularize Eq.~(\ref{eq:Cl22}) by integrating from some finite minimum
distance corresponding to the typical distance to the nearest source, $r_{\rm min}\sim1\,$Mpc. Furthermore,
Eq.~(\ref{eq:Cl22}) is also formally regularized at $z_{\rm min}\to0$ by the spatial extent of nearby
sources, described by $|u(k,{\cal P})|^2$. For the NFW profile, the mass of the halo within
distance $r$ from the halo centre increases as $r^2$ up to $r=r_s$, and then increases logarithmically
between $r_s$ and $r_v$ since $\rho_h(r)\propto r^{-3}$, see Eq.~(\ref{eq:NFW}). Therefore,
the dominant contribution to the halo mass comes from $r<r_s$. Similarly, for $r<r_s$ the annihilation
signal increases as $r$, but between $r_s$ and $r_v$ increases only as $r_s^{-3}-r^{-3}$. Assuming the
emission traces $\rho_h$ for astrophysical emission processes and $\rho_h^2$ for dark
matter annihilation, the Fourier transforms of these dependencies then give
$u(k,{\cal P})\propto k^{-\gamma}$ for $k\gg r_s^{-1}$, with $\gamma=2$ for astrophysical emission
and $\gamma=1$ for dark matter, see Appendix A.6 for more details. Since $k=l/r(z)$, and thus
$|u\left[l/r(z),{\cal P}\right]|^2\propto r(z)^{2\gamma}$, the one-halo term in Eq.~(\ref{eq:Cl22})
diverges only for $\gamma\le0.5$. Therefore, under our assumptions for the emission profile,
Eq.~(\ref{eq:Cl22}) is convergent even without cut-offs in either $r_{\rm min}$ or the apparent
luminosity. Since nevertheless in particular the one-halo term is quite sensitive to nearby sources,
in the following we study  its dependence on $S_{\rm cut}$ and $r_{\rm min}$.

\begin{figure}
\includegraphics[width=0.6\textwidth]{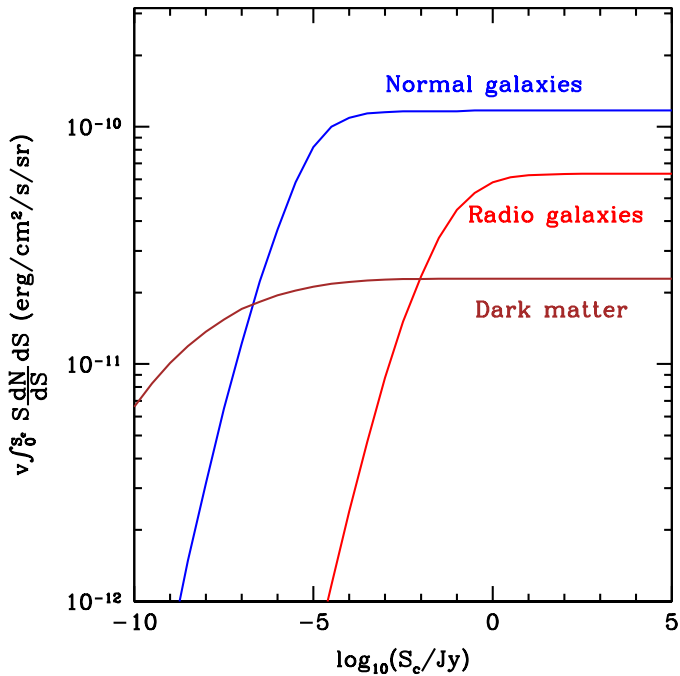} 
\caption{The cumulative contribution of sources of apparent luminosity $S$ smaller than $S_{\rm cut}$ to the
two-halo term at $2\,$GHz. The red, blue and brown lines represent the contribution from radio galaxies,
normal galaxies, and dark matter (fiducial scenario with $Y_e=10$), respectively.}
\label{fig:sc_2G}
\end{figure}  

Fig.~\ref{fig:sc_2G} shows the dependence of the two-halo term on $S_{\rm cut}$. According to
Eq.~(\ref{eq:L22}), the two-halo term scales with the square of the average flux. Since the
apparent luminosity of radio galaxies can be of the order of a Jansky
($1\,{\rm Jy}=10^{-23}\,{\rm erg}\,{\rm cm}^{-2}\,{\rm Hz}^{-1}\,{\rm s}^{-1}$), the two-halo term from
radio galaxies starts to decline
when we cut sources being less luminous than a critical luminosity below a Jansky. In contrast, the
contribution of normal galaxies which are much less luminous than radio galaxies starts to decline
only when we cut sources more luminous than $\simeq10\mu$Jy. The contributions of dark matter halos
to the dark matter annihilation signal is basically unaffected by any source removal, even to luminosities
down to $\sim1\mu\,$Jy. This is easy to explain: In our fiducial scenario the largest dark halos of about $10^{14}\,M_\odot$
produce only about $1.3\times10^{38}\,{\rm erg}/$s  at 2 GHz from dark matter annihilation, far less than
the typical radio luminosity of galaxies of about $2\times10^{40}\,{\rm erg}$/s. As a result, removing bright
sources increases the contribution of dark matter annihilation to the two-halo term relative to the
contribution from astrophysical sources.

\begin{figure}
\includegraphics[width=0.6\textwidth]{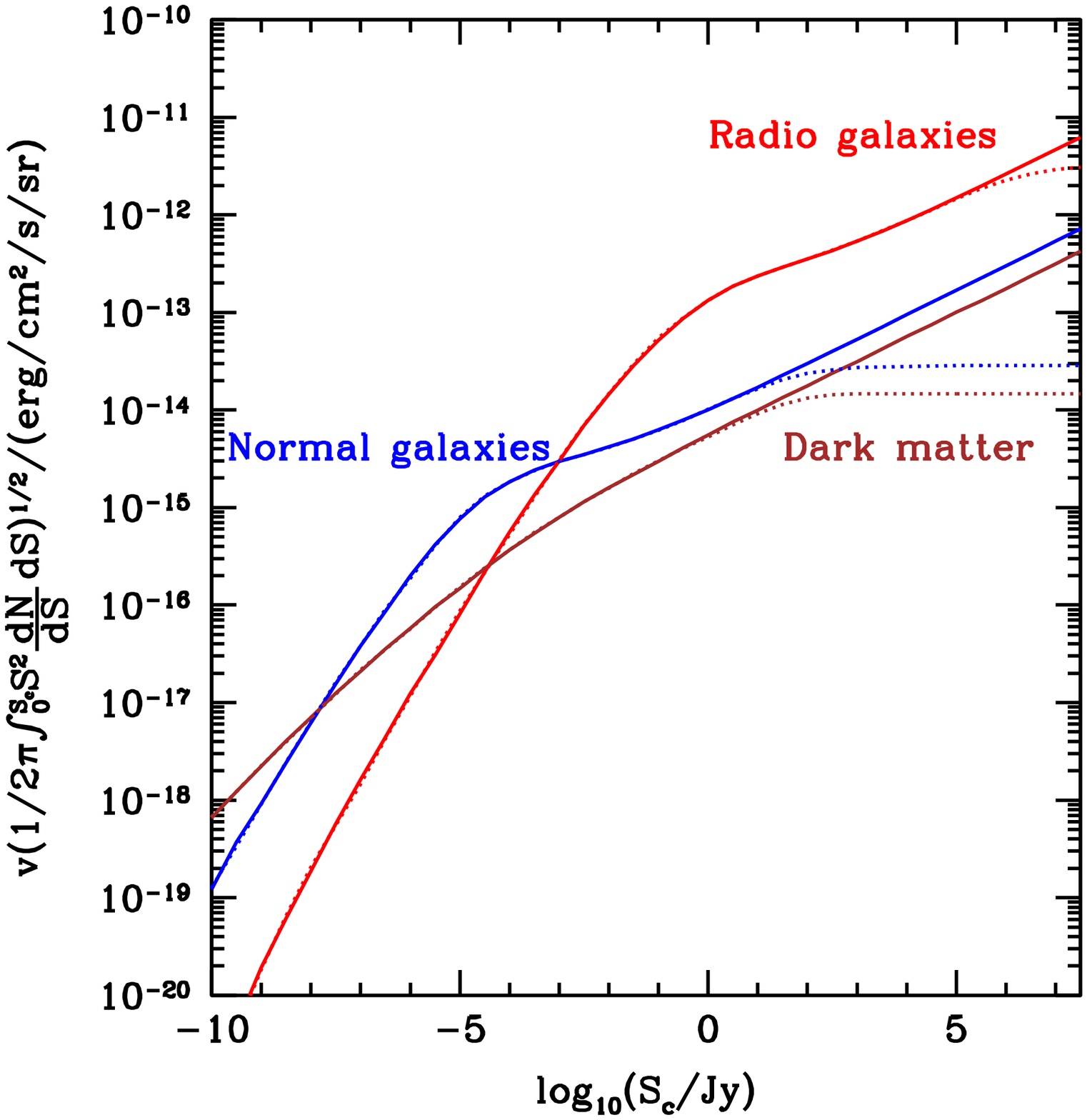}
\caption{The cumulative contribution of sources of apparent luminosity $S$ smaller than $S_{\rm cut}$
to the one-halo (Poisson) term at 2 GHz. The solid and dotted curve represent the cases of 
$r_{\rm min}= 0$ and $r_{\rm min}= 1$ Mpc, respectively. Color keys are as in Fig.~\ref{fig:sc_2G}.}  
\label{fig:ssc_2G}
\end{figure} 

Next we discuss the one-halo term. The one-halo term is more sensitive to the cut-offs in apparent
luminosity $S_{\rm cut}$ and to the minimal distance $r_{\rm min}$ than the two-halo
term because of two reasons: First, the two-halo term Eq.~(\ref{eq:L22}) is the square of an integral
of luminosities, whereas the one-halo term Eq.~(\ref{eq:L21}) is essentially Poisson noise and thus
proportional to an integral of squared luminosities, which makes the contribution from bright
sources more important. Second, the two-halo term is further regularized by
the linear power spectrum at large $k=l/r(z)$. In
Fig.~\ref{fig:ssc_2G} we show the cumulative contribution of sources dimmer than $S_{\rm cut}$
to $C_l^{1h}$. Similarly to the two-halo term shown in Fig.~\ref{fig:sc_2G}, the contribution
of radio galaxies and ordinary galaxies decreases rapidly below $\simeq1\,$Jy and $10\,\mu$Jy,
respectively, whereas the contribution of dark matter annihilation is affected less by source removal.
Nevertheless, the contribution of bright sources is now much larger than for the two-halo term,
as expected, and the one-halo term continues to rise with inclusion of brighter sources.
On the other hand, practically one should cut off the integral at some minimal distance
$r_{\rm min}\simeq1\,$Mpc within which there are essentially no bright sources. Since
sources at small distance appear bright, the cut-off in luminosity and minimal distance is of
course to some extent degenerate, as confirmed by Fig.~\ref{fig:ssc_2G}. For radio galaxies,
removal above $\simeq10^5\,$Jy is equivalent to restricting to distances larger than 1 Mpc.
For ordinary galaxies, cutting at a minimal distance $r_{\rm min}=1\,$Mpc is equivalent to
removing sources brighter than 0.1 Jy. Since observational sensitivities are considerably
better than these luminosities, cutting at $r_{\rm min}\simeq1\,$Mpc does, therefore, not introduce any
significant uncertainties. Note that in Fig.~\ref{fig:sc_2G} dark matter dominates the
two-halo terms if all sources above $\simeq0.1\mu\,$Jy are removed, while in Fig.~\ref{fig:ssc_2G} it would
dominate the one-halo terms only for unrealistically small cut-off luminosities $\la1\,$nJy.
This is because the one-halo term is much more sensitive to bright sources than the two-halo
term and because the dark matter contribution consists of dimmer sources than ordinary
astrophysical sources.

\begin{figure}
\includegraphics[width=0.6\textwidth]{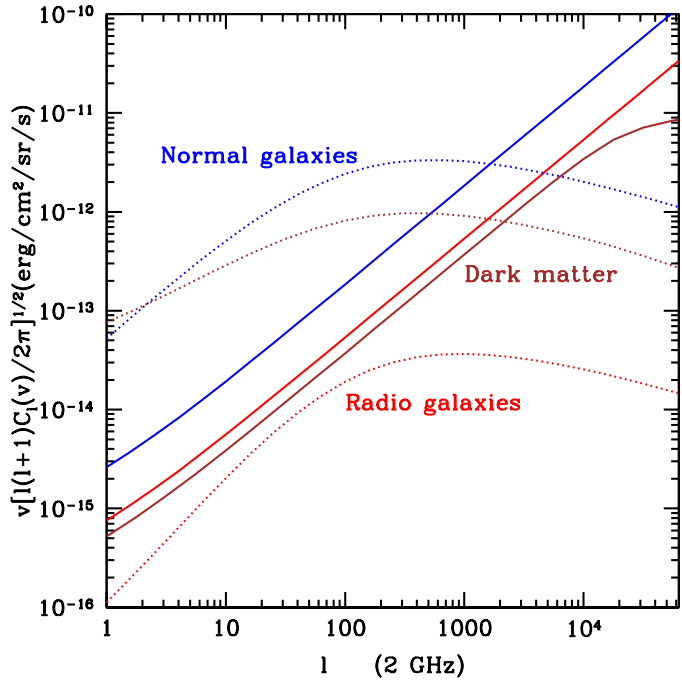}
\caption{Angular power spectra of various components at $2\,$GHz. Solid lines and dotted lines
represent the one-halo and two-halo terms, respectively. We assume the astrophysical sources
to be point-like. The minimal dark matter halo mass is $M_{\rm min}= 10^{6}\,M_{\odot}$.
Sources at distances below $r_{\rm min}=1\,$Mpc, and of apparent luminosity
above $S_{\rm cut}=0.1\,$mJy were removed. Color keys are as in Fig.~\ref{fig:sc_2G}.}
\label{fig:r1-point-sim-01mj}
\end{figure}

\begin{figure}
\includegraphics[width=0.6\textwidth]{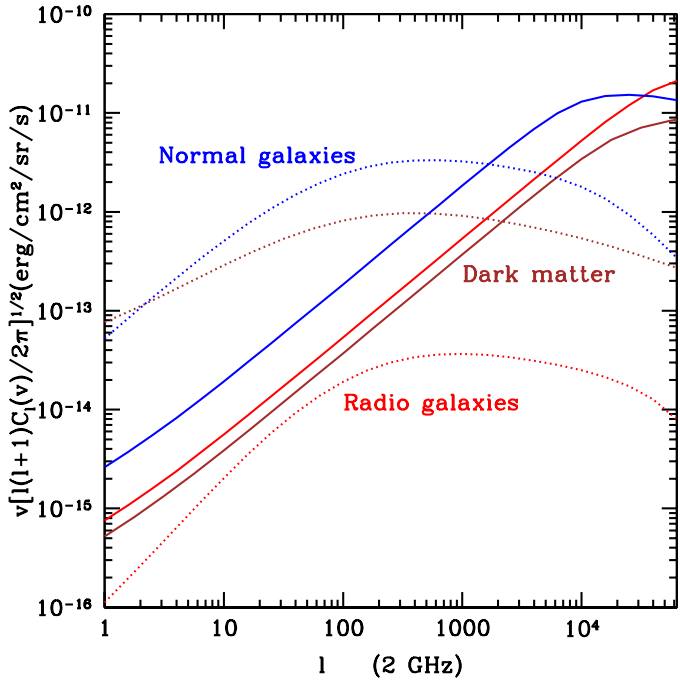}
\caption{Same as Fig.~\ref{fig:r1-point-sim-01mj}, but assuming the emission profile of the astrophysical
sources follows an NFW profile. Sources with luminosities above $S_{\rm cut}=0.1\,$mJy are again
subtracted.}
\label{fig:z0-nfw-sim-01mj}
\end{figure}

\begin{figure}
\includegraphics[width=0.6\textwidth]{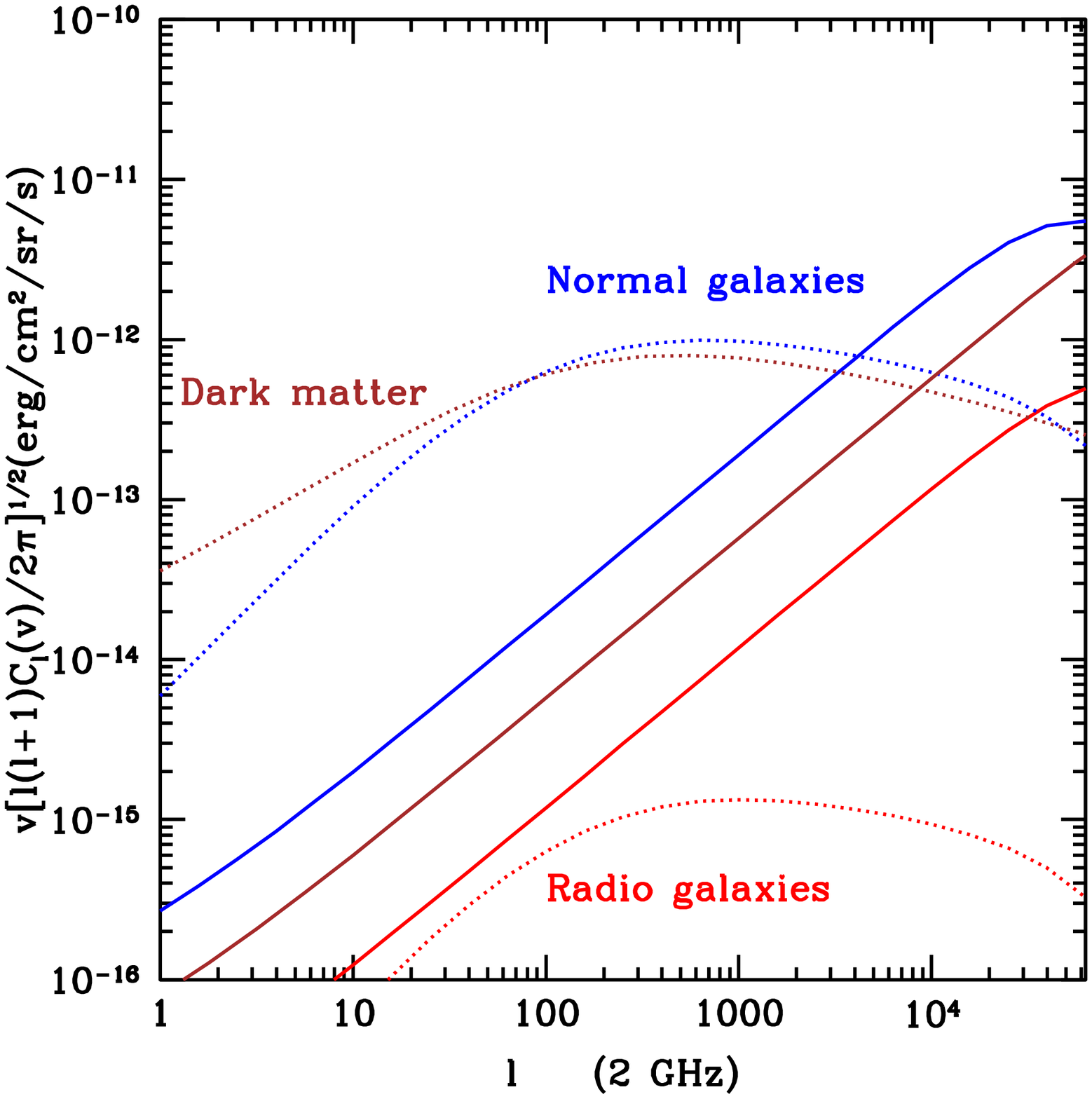}
\caption{Same as Fig.~\ref{fig:z0-nfw-sim-01mj}, but subtracting sources above $S_{\rm cut}=1\,\mu$Jy.}
\label{fig:z0-point-sim-1uj}
\end{figure}

The angular power spectra of the radio background at 2 GHz produced by galaxies and by our fiducial dark matter
scenario are shown in Figs.~\ref{fig:r1-point-sim-01mj},~\ref{fig:z0-nfw-sim-01mj}, and~\ref{fig:z0-point-sim-1uj}
for different source removal cuts. Based on the above discussion, the qualitative behavior of the one-
and two-halo terms can be easily understood: In Fig.~\ref{fig:r1-point-sim-01mj} we assume galaxies to
appear point-like and we remove sources brighter than $0.1\,$mJy. The one-halo terms from these sources
thus increase proportional to $\left[l(l+1)\right]^{1/2}$ in the above figures. The same applies
to the one-halo term of the dark matter contribution
for $l\la10^4$, corresponding to angular scales $\theta\simeq\pi/l\ga0.02^\circ$. At smaller
angular scales the power spectrum is suppressed by the inner structure of the dark matter halos. We can
estimate this critical scale as follows: The one-halo term is dominated by the brightest halos which
correspond to the largest and nearest halos. In our fiducial scenario, the annihilating dark matter in
the largest halos can emit a radio flux of $\sim7\times10^{28}\,{\rm erg}\,{\rm s}^{-1}\,{\rm Hz}^{-1}$ at 2 GHz,
such that the minimum co-moving distance is $r\simeq830\,$Mpc, $z\simeq0.2$ if sources brighter than
$0.1\,$mJy are removed. The scale $r_s$ for the corresponding $10^{14}\,M_{\odot}$ halo is about 0.21 Mpc.
This corresponds to a multipole $l\simeq\pi r/r_s\simeq1.2\times10^4$. This simple estimation is consistent
with our detailed calculation shown in Fig.~\ref{fig:r1-point-sim-01mj}.

Figs.~\ref{fig:r1-point-sim-01mj},~\ref{fig:z0-nfw-sim-01mj}, and~\ref{fig:z0-point-sim-1uj} show that
for radio galaxies the one-halo term is always larger than the two-halo term at all multipoles, as
expected because of the high luminosity of radio galaxies. For dark matter and normal galaxies, the
two-halo term dominates at small $l$. The dependence of the angular power spectrum on $l$ can potentially
be used to discriminate the dark matter signal from astrophysical contributions: For $l\la3\times10^3$, the
annihilation power spectrum looks significantly flatter than the signal from normal galaxies. In other words,
at large angular scales, the annihilation signal has relatively more power. This can be understood
as follows: After
cutting bright sources, many more dim nearby annihilation sources than galaxies contribute. In addition,
at large redshift the synchrotron emission from dark matter annihilation is suppressed by the increased
inverse Compton scattering rate on the CMB, see Eq.~(\ref{eq:I-syn-dm2-w}). The two-halo term is proportional to
$P_{\rm lin}(k)$ which peaks at $\simeq0.03\,{\rm Mpc}^{-1}$, corresponding to $l\simeq0.03\,r(z)/$Mpc.
The on average smaller distance to the dark matter halos then translates into relatively more
power at small $l$.

In Fig.~\ref{fig:z0-nfw-sim-01mj}, we take into account the spatial extent of the radio emission
of galaxies. We assume the luminosity profiles of galaxies to be roughly proportional to the dark matter density
profile which is obtained following Appendix A.2 with the halo mass is obtained from the relation
between mass and bolometric luminosity~\cite{Spinoglio:1995pg}.
As a result, for normal and radio galaxies the one-halo term
starts to drop for $l\ga6000$ and $l\ga2.5\times10^4$, respectively. Compared to the dark matter signal,
the suppression thus sets in at slightly smaller $l$ for normal galaxies, but only at larger $l$
for radio galaxies. For normal galaxies this is due to the more extended emission profile which
more closely follows the density as opposed to the squared density in case of dark matter. This is also reflected
by the Fourier transform of the emission profiles shown in Fig.~\ref{fig:fourier_rho}. For radio galaxies
this effect is overcompensated by the fact that they are much brighter such that after cutting bright
nearby sources, their average distance is much larger where their angular extent appears smaller.

Since future radio detectors such as the
square kilometer array (SKA)~\cite{Jackson:2004vw} can reach point flux sensitivities of $\sim1\mu$Jy,
we show the power spectra of the background remaining after a corresponding luminosity cut
in Fig.~\ref{fig:z0-point-sim-1uj}. Since the two-halo term from dark matter annihilation is insensitive
to such luminosity cuts whereas the contribution from galaxies decreases rapidly,
as shown in Fig.~\ref{fig:sc_2G}, the relative contribution of dark matter annihilation increases
and gives rise to a flatter power spectrum at moderate $l$.

\begin{figure}
\includegraphics[width=0.6\textwidth]{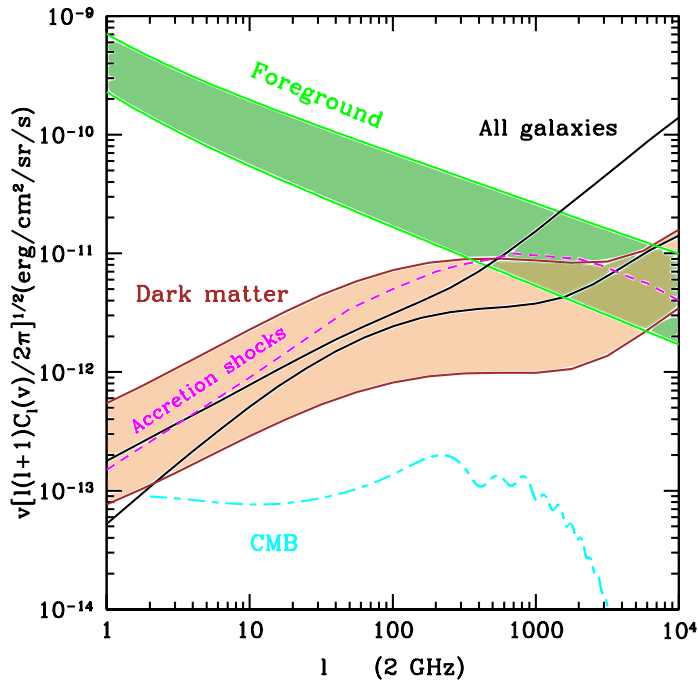}
\caption{Angular power spectra of the radio sky at 2 GHz compared with various estimates of the
Galactic foreground at Galactic latitude $b>20^\circ$ (green shaded region) and the CMB (cyan curve). The
brown band represents the annihilation spectrum, where the upper and lower ends
correspond to $F_{\rm dm}=10$ and $F_{\rm dm}=1$, respectively, see Eq.~(\ref{eq:F_dm}),
and from which halos brighter than $0.1\,$mJy were removed.
The black-dotted and black-solid curves represent the total signal from normal and radiogalaxies,
for luminosity cuts $S_{\rm cut}=10\,$mJy and $S_{\rm cut}=0.1\,$mJy, respectively. Also shown is
a possible contribution from intergalactic shocks~\cite{Keshet:2004dr}, normalized such that
its angular power spectrum is comparable to the Galactic foreground.}
\label{fig:cl_sum}
\end{figure}

We now have to compare the cosmological background power spectra discussed so far with other potential contaminations. Fig.~\ref{fig:cl_sum} compares the signals from ordinary and
radio galaxies and from our fiducial dark matter scenario with the power spectra of the
CMB and of the Galactic foreground at high Galactic latitude. The power spectrum of the Galactic
foreground is not very well measured and we represent its uncertainties as a green band in Fig.~\ref{fig:cl_sum}.
At high Galactic latitude below 10 GHz local
foreground fluctuations dominate over the CMB power spectrum which is why the CMB anisotropy
measurements are performed above 20 GHz. Concerning annihilation signatures of dark matter with
mass $m_X\ga100\,$GeV in the angular power spectrum of the radio sky, the optimal frequency band is
around 2 GHz. At higher frequencies, the synchrotron emission of electrons produced from
dark matter annihilations cuts off due to Eq.~(\ref{eq:E_crit}) and the CMB signal increases. At lower frequencies,
synchrotron emission by Galactic electrons dominates the power spectrum even at high Galactic latitude~\cite{Tegmark:1995pn,Tegmark:1999ke,Giardino:2001jv,LaPorta:2008ag}. Around 2 GHz,
Galactic synchrotron emission always dominates, whereas free-free emission is a factor few smaller.

Also shown in Fig.~\ref{fig:cl_sum} is a possible signal from intergalactic shocks~\cite{Keshet:2004dr}.
Since its normalization is rather uncertain, we normalized it such that it is comparable to the average
estimate of the Galactic foreground. The thermal SZ effect~\cite{Sunyaev:1970eu} is another characteristic contamination caused by hot ionized gas in galaxy clusters and filaments outside of clusters~\cite{persi:1995}.
Since it dominates at small angular scales, $l\ga3000$, and at high frequencies above 30 GHz, we can neglect this effect here.

As can be seen from comparing Fig.~\ref{fig:z0-nfw-sim-01mj} and~\ref{fig:z0-point-sim-1uj} and from 
Fig.~\ref{fig:cl_sum}, future radio telescope arrays sensitive around $\nu\sim2\,$GHz, with
their higher point flux sensitivities should allow to further reduce the contribution from galaxies,
whereas for $l\la6000$ the dark matter contribution is hardly changed by removing still fainter
sources. This can be understood from the fact that the dark matter signal is dominated by the two-halo
term which is insensitive to $S_{\rm cut}$ for $S_{\rm cut}\ga1\,$nJy, see Fig.~\ref{fig:sc_2G}.
This shows that for dark matter annihilation the distribution of $l(l+1)C_l$ is nearly
flat for $200\la l\la2000$. At smaller $l$ the power spectrum is dominated by
Galactic foregrounds and at larger $l$ the one-halo term from galaxies grows rapidly.
The most sensitive range $200\la l\la3000$ should be accessible to present and
future radio telescopes with their high angular resolution. The SKA will have a sensitivity
of about $6\times10^{-13}\,{\rm erg}\,{\rm cm}^{-2}\,{\rm sr}^{-1}\,{\rm s}^{-1}$ in
the units of the above figures.

We conclude that the power spectrum from dark matter annihilation tends to be flatter than other
contributions because of an interplay of the following effects:

\begin{itemize}

\item The astrophysical signals are dominated by fewer and much brighter sources than the dark matter annihilation signal which consists of many faint sources. For $S_{\rm cut}=0.1\,$mJy, the two-halo term
dominates for $l\la10^3$ for both the signals
from galaxies and from dark matter annihilation. In addition, the dark matter signal is significantly
flatter in that angular range, i.e. it has relatively more power at small $l$. This is because the
two-halo term is proportional to the linear power spectrum whose peak in wavenumber for the on average closer
and dimmer dark matter annihilation sources translates into smaller $l$ at these luminosities.

\item For $l\ga10^4$, the inner spatial structure of the galaxies and dark matter halos
becomes important. The inner structure tends to suppress the power spectra, but the exact angular scale at which these effects become important depends on the halo size, the profile of
the emission and source luminosity cut-off.

\item The various components evolve differently with the Universe expansion.
For example, at high redshift inverse Compton scattering on the CMB tends to suppress synchrotron
emission in dark matter halos, whereas astrophysical sources such as radio galaxies tend to
be more active at $z\simeq3$.

\end{itemize}

\subsection{Dark Matter Constraints}
We can now scale the dark matter signal to parameter values different from the fiducial
scenario, by multiplying with the factor
\begin{equation}\label{eq:F_dm}
  F_{\rm dm}\equiv\left(\frac{A_b}{10}\right) \left(\frac{Y_e}{10}\right)
  \left(\frac{\left\langle\sigma v\right\rangle}{3\times10^{-26}{\rm cm}^3{\rm s}^{-1}}\right)
  \left(\frac{100 {\rm GeV}}{m_X}\right)^2\left(\frac{10\,\mu{\rm G}}{B}\right)^{1/2} 
\end{equation}
We caution that a boost factor as high as $A_b\simeq10$ has not been verified in all dark matter
structure simulations and that the average magnetic field $B$ could be significantly smaller than
10$\mu$G if many small-scale subhalos contribute. However, smaller values for $A_b$ and $B$ partially
compensate in Eq.~(\ref{eq:F_dm}) so that one could still obtain observable signatures as long as $B$ is
large enough to produce emission at GHz frequencies, see Eq.~(\ref{eq:E_crit}).

If we choose $F_{\rm dm}=10$ (upper end of brown band in Fig.~\ref{fig:cl_sum}) and
$S_{\rm cut}=0.1\,$mJy, the annihilation
spectrum dominates over other cosmological backgrounds for $100\la l\la10^4$ and should become
distinguishable from the Galactic foreground. Note that this foreground is likely further reduced
close to the Galactic poles. In this situation it should thus be
possible to disentangle the rather flat power spectrum of the dark matter annihilation signal
from other contributions in the range of $200\la l\la3000$. We can thus assert that radio observations
are sensitive to
\begin{equation}\label{eq:limit}
  F_{\rm dm}\ga10\,,
\end{equation}
with some dependence on the source luminosity cut-off $S_{\rm cut}$.
Note that the dark matter signal shown in Fig.~\ref{fig:cl_sum} does not strictly scale
with $F_{\rm dm}$ because it depends on sources dimmer than $S_{\rm cut}$, here chosen as 0.1 mJy.
However, since few brighter dark matter halos contribute, the signal scales with
$F_{\rm dm}$ in first approximation.

\begin{figure}
\includegraphics[width=0.6\textwidth]{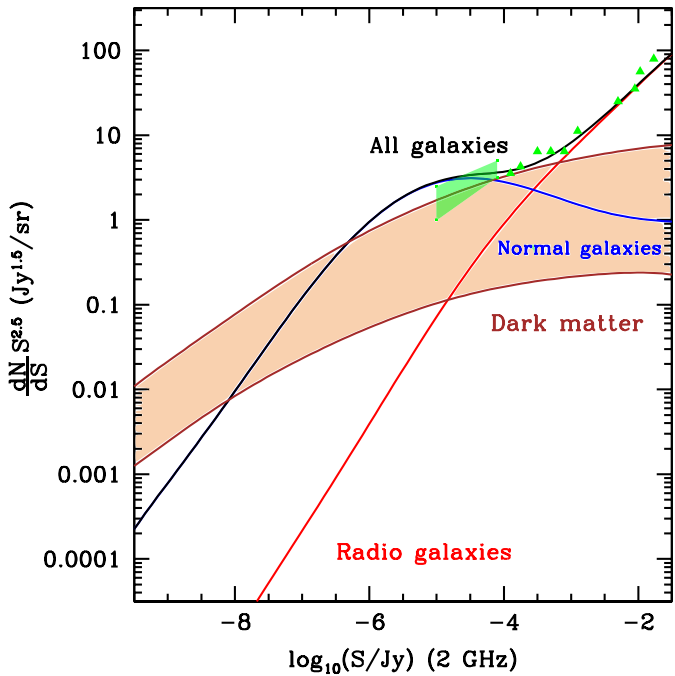}
\caption{Observed radio source counts $(dN/dS)S^{2.5}$ as function of apparent radio flux $S$
compared with predictions for normal galaxies (blue curve), radio galaxies (red curve), and
annihilations from dark matter halos (brown band, for $1\la F_{\rm dm}\la10$). Green shaded region
and triangles are data from Ref.~\cite{condon}.}
\label{fig:counts}
\end{figure}

One can also compare observed radio source counts as a function of apparent point source flux with
predictions for astrophysical sources and dark matter annihilation sources. This is done
in Fig.~\ref{fig:counts} for the same parameters as used in Fig.~\ref{fig:cl_sum}. This
establishes the constraint $F_{\rm dm}\la10$.
In contrast, Fig.~\ref{fig:cl_sum} provides dark matter signatures for future measurements but
currently does not allow to put a constraint on $F_{\rm dm}$ because of the uncertainties in the Galactic
foreground spectrum. Note that a future observational extension of the source count spectrum
in Fig.~\ref{fig:counts} to apparent luminosities $S\la\,\mu$Jy will provide an additional test
for dark matter which predicts a shallower source count distribution than astrophysical sources.

We have not computed the contribution from dark matter annihilations in our own Galaxy
to the anisotropic radio flux in the present work. However, we know from
Refs.~\cite{Hooper:2007gi,Grajek:2008jb} that
for our fiducial values for cross section and mass, at least the smooth halo component does
not lead to fluxes higher than current observations from WMAP. The contribution from
Galactic substructures is probably more model dependent than our cosmological flux which apart
from an overall boost factor depends
only on the host halo distribution and effectively averages over a much larger ensemble of
halos. This can also be seen from Ref.~\cite{SiegalGaskins:2008ge} where the predictions of the $\gamma-$ray flux
from Galactic dark matter annihilations varied over orders of magnitude.

\section{Conclusions}

Many different indirect detection signatures have been investigated to constrain the parameter space
for dark matter~\cite{Ando:2005xg,Ando:2006mt,Chen:2003gz,Zhang:2006fr,Zhang:2007zzh,Myers:2007fj,Padmanabhan:2005es}.
In the present paper we have calculated intensity and angular power spectrum of the cosmological
background of synchrotron emission from the electrons and positrons produced in annihilations of cold dark matter.
The resulting radio background around $\simeq2\,$GHz and its angular power spectrum for multipoles
$200\la l\la3000$ has comparable or better sensitivity to dark matter annihilation cross sections
than other signatures. Furthermore, a comparison of observed radio source counts with predictions
for dark matter annihilation results in the constraint $F_{\rm dm}\la10$ for the parameter
defined in Eq.~(\ref{eq:F_dm}). Under reasonable assumptions on dark matter clustering and magnetic fields
in the halo environment, the range of annihilation cross sections corresponding to the constraint
$F_{\rm dm}\la10$ is comparable to constraints from synchrotron emission in an NFW 
profile~\cite{Hooper:2007gi,Grajek:2008jb}. Galactic $\gamma-$ray constraints derived under similar
assumptions~\cite{Dodelson:2007gd} are also comparable. The sensitivity of our signal is
considerably better than conservative limits based on annihilation into neutrinos~\cite{Yuksel:2007ac},
and comparable to limits on annihilation into $\gamma-$rays from diffuse cosmological emission~\cite{Mack:2008wu}.

Sensitivities to values of order ten for the parameter $F_{dm}$ defined in Eq.~(\ref{eq:F_dm})
are interesting for
non-thermal dark matter whose annihilation cross sections can be larger than our fiducial value
$\left\langle\sigma v\right\rangle=3\times10^{-26}\,{\rm cm}^3\,{\rm s}^{-1}$, the
cross section required for thermal dark matter. An example for a non-thermal dark matter candidate
with large cross section is the wino LSP occurring in supersymmetric theories with anomaly
mediation~\cite{Randall:1998uk,Giudice:1998xp}. We believe that radio observations in particular
with future instruments such as SKA can provide valuable information on dark matter.

\section*{Acknowledgements}
We acknowledge partial support by the DFG (Germany) under grants SFB-676 and
GK 602, and by the European Union under the ILIAS project (contract No.\ RII3-CT-2004-506222).
L.Z. would like to thank Shin'ichiro Ando, Yanchuan Cai and Yan Qu for valuable discussions.

\appendix
\section{Cosmological Dark Matter Distribution}

\subsection{Power Spectrum and Halo Mass Function}

The mass function of the halo distribution is derived from the Press-Schechter formalism~\cite{Press:1973iz}.
In this approach fluctuations in the linear density field with
$\delta > \delta_{\rm c}$ decouple from the local Hubble expansion of the universe and collapse to
form non-linear structures. The fraction of the volume that has collapsed is predicted to be
\begin{equation}
f_{\rm coll} (M(R), z) = \frac{2}{\sqrt{2 \pi} \sigma(R,z)} \int_{\delta_{\rm c}}^{\infty} d \delta \, e^{-\delta^2/2\sigma^2(R,z)}\, ,
\end{equation}
where $R$ is the co-moving radius over which the density field has been smoothed, which is related to the halo mass by
$M(R) = \rho_m4\pi R^3/3$ with $\rho_m$ the co-moving matter density of the universe.
The number density of halos is then found to be given by~\cite{Press:1973iz,Sheth:1999mn}
\begin{equation}
\frac{dn(M,z)}{dM} = -\frac{\rho_{m}}{M} \frac{d f_{\rm coll}(M(R),z)}{dM} =
\frac{\rho_m}{M}f(\nu)\frac{d\nu}{dM}\,,
\label{eqn:massfunc}
\end{equation}
where
\begin{equation}\label{equ:fnu}
f(\nu) \equiv \sqrt{\frac{2A^2a^{2}}{\pi}}[1 + (a\nu^2)^{-p}]e^{-\frac{a\nu^2}{2}} \,.
\label{eqn:fnu}
\end{equation}
Here
\begin{equation}
\nu(M,z) \equiv \frac{\delta_{\rm c}(z)}{\sigma(M,z)} \, ,
\end{equation}
and
\begin{equation}
\delta_{\rm c}(z) \simeq 1.686
\end{equation} is the critical density required for spherical
collapse at a redshift $z$ in an Einstein-de Sitter space.
The variance in the density field smoothed with a top-hat filter of radius
$R=({3M}/{4\pi \rho_m})^{\frac{1}{3}}$ is
\begin{equation}
\sigma^2(M,z) = G^2(z)\int \frac{dk}{k} \frac{k^3 P_{\rm lin}(k)}{2\pi^2} |W(kR)|^2\,,
\end{equation}
where
\begin{equation}
W(x)=\frac{3}{x^3}[{\rm sin}(x)- x{\rm cos}(x)]\,,
\end{equation}
$P_{\rm lin}(k)$ is the linear matter power spectrum, and
\begin{equation}\label{equ:growth}
G(z) =\frac{H(z)\int_{z}^{\infty}dz^{\prime}(1+z^{\prime})[H(z^{\prime})]^{-3}}
{H_0\int_{0}^{\infty}dz^{\prime}(1+z^{\prime})[H(z^{\prime})]^{-3}}
\end{equation}
is the growth factor with linear perturbation theory, often also denoted by $D(z)$.
In Eq.~(\ref{equ:fnu}) $A$, $p$, and $a$  are constants, with
the canonical Press-Schechter (PS) and Sheth-Tormen (ST) mass functions corresponding
to the parameters $(p=0,a=1)$ and $(p=0.3,a=0.707)$, respectively. The normalization $A$
is determined by requiring mass conservation such that
\begin{equation}
\frac{1}{\rho_m}\int_0^{\infty}dM M \frac{dn}{dM} = \int_0^{\infty}d\nu f(\nu)=1 \,.
\end{equation}
For PS $A=1$ and for ST $A=0.3222$.

The primordial power spectrum $P(k)\propto A_sk^{n_s}$ can be modified by the content and evolution of
different matter components of the Universe due to the perturbations that enter the horizon at
different epochs. This allows one to relate the linear
power spectrum to the primordial power spectrum through a transfer function $T(k)$ via
\begin{equation}
 P_{\rm lin}(k,z)=D^2(z)P_{\rm lin}(k,z=0)=D^2(z)A_s (k\cdot{\rm Mpc})^{n_s}T^2(k) 
\end{equation}
Fitting formula for an adiabatic CDM model give~\cite{Bardeen:1985tr} 
\begin{equation}
T_{\rm CDM}(q)=\frac{\ln(1+2.34q)}{2.34q}[1+3.89q+(16.1q)^2+(5.46q)^3+(6.71q)^4]^{-1/4} 
\end{equation}
where $q=k\cdot{\rm Mpc}/(h\Gamma)$ and $\Gamma=\Omega_m h\exp[\Omega_b(1+\sqrt(2h)/\Omega_m)]$.
One usually uses the rms fluctuation on an $8h^{-1}\,$Mpc scale to normalize the amplitude of the present power spectrum. From WMAP 5 year data, we adopt $n_s=0.96$~\cite{Komatsu:2008hk,Dunkley:2008ie}, and
$A_s=1.4\times10^7$. Following Ref.~\cite{Carroll:1991mt}, we may furthermore write the linear growth
factor as
\begin{equation}\label{Dz}
D(z)=\frac{1}{1+z}\frac{g(z)}{g(0)} \,,
\end{equation}
where an approximate expressionfor $g(z)$ is
\begin{equation}
g(z)= \frac{5/2\Omega_m(z)}{\Omega_m(z)^{4/7}-\Omega_\Lambda(z)+(1+\Omega_m(z)/2)(1+\Omega_\Lambda(z)/70)}\,.
\end{equation}
 
\subsection{Dark Matter Density Profile}
The  halo mass function has to be supplemented by the dark matter density profile. For the dark matter
profile within each halo we use an NFW profile~\cite{Navarro:1995iw},
\begin{equation}\label{eq:rho_NFW}
\rho_h({\bf r}) = \frac{\delta_{ch}\rho_m(1+z)^3}{r/r_s(1+r/r_s)^2}\,,
\end{equation}
where $r_s$ is a characteristic radius.
Within the context of the spherical collapse model, the outer extent of the cluster is taken to be
the virial radius
\begin{equation}
r_v=\left[\frac{3M}{4\pi\rho_m(1+z)^3\Delta_c(z)}\right]^{\frac{1}{3}} \, ,
\end{equation}
where $\rho_m(1+z)^3$ is the average physical background matter density of the universe at redshift $z$, and
\begin{equation}
\Delta_c(z) \simeq 18\pi^2\left[ 1+\frac{88}{215} \left(\frac{ 1 - \Omega_m}{\Omega_m(1+z)^3}\right)^{\frac{86}{95}} \right]
\end{equation}
is the overdensity of the halo relative to the
background density~\cite{Henry:2000bt}.
The ratio of the virial radius to the scale radius is called the concentration parameter $c\equiv{r_v}/{r_s}$.
A combination of the definitions of virial mass and density profile gives
\begin{equation}\label{eq:delta_ch}
\delta_{ch}=\frac{\Delta_c(z)}{3}\frac{c^3}{\ln(1+c)-c/(1+c)}
\end{equation}
Together, $c$ and $M$ completely determine the dark matter distribution of a given halo.
Note that an NFW profile is a conservative assumption compared to steeper profiles that
have been proposed, for example the Moore profile~\cite{Moore:1999gc}, which would consequently
lead to a larger dark matter signal.

\subsection{Concentration Distribution}
It is still uncertain how the concentration parameter depends on mass and redshift. In our paper
we consider a realistic value of $c(M=10^{-6}\,M_\odot,z=0)\simeq70$. Extrapolating to the low mass range, we
use~\cite{Bullock:1999he}
\begin{equation}
c(M,z)=4\,\frac{1+z_c}{1+z}\,,
\end{equation} 
where the collapse redshift $z_c$ is implicitly given by the relation $M_*(z_c)=0.01M$,
where $M_*(z)$ is the mass scale at which $\sigma(M_*,z)=\delta_c$. In a less conservative
parametrization motivated by numerical simulations $c$ would follow a log-normal
distribution with standard deviation $\sigma_c=0.18$,
\begin{equation}\label{eqn:pc}
{\cal P}(\ln c|M,z) =\frac{1}{\sqrt{2\pi}\sigma_c}\exp\left({{-\frac{[\ln c- \ln{\bar{c}(M,z)}]^2}{2\sigma_c^2}}}\right)\,,
\end{equation}
where the mean concentration parameter $\bar{c}$ is related to the halo mass via~\cite{Cooray:2002dia}
\begin{equation}\label{eqn:pc2}
\bar{c}(M,z)=\frac{c_0}{1+z}\left[ \frac{M}{M_{*}(z=0)} \right]^{-\alpha_{c}}\,,
\end{equation}
where $c_0$ and $\alpha_c$ are constants whose numerical values~\cite{Bullock:1999he} are
typically chosen to be $c_0=9$ and $\alpha_c=0.13$. However, application of
this  parameterization to low-mass halos and to high redshift give values inconsistent
with some simulations~\cite{Diemand:2005vz}. A third parameterization~\cite{Seljak:2000gq}
uses
\begin{equation}\label{eqn:sj}
\bar{c}(M,z)=a(z)\left[ \frac{M}{M_{*}(z)} \right]^{b(z)}\,,
\end{equation}
with $a(z)=10.3(1+z)^{-0.3}$ , and $b(z)=0.24(1+z)^{-0.3}$. The parametrization Eqs.~(\ref{eqn:pc}),
(\ref{eqn:pc2}) give comparable dark matter signals, whereas the parametrization Eqs.~(\ref{eqn:sj})
would lead to signals about a factor 20 higher than our conservative calculation.

\subsection{Bias}
For the linear dark matter halo bias $b(M,z)$ appearing in Eq.~(\ref{eq:L22}) we adopt~\cite{Mo:1996zb} 
\begin{equation}
b(M,z)=1+\frac{\nu^2(M,z)}{\delta_cD(z)} \,,
\end{equation}
whereas for the galaxy bias we simply use unity.

\subsection{Bolometric Luminosity}
In order to compute the Fourier transform of galaxy density profiles, we first need the galaxy
mass. For normal galaxies we use the relation~\cite{Spinoglio:1995pg}
\begin{equation}
M=14\,\frac{L_{60}\nu_{60}}{L_{sun}}\frac{\Omega_m}{\Omega_b}\,M_\odot\,,
\end{equation}
where $L_{60}$ is the luminosity at 60 microns, and $\nu_{60}\sim5000\,$ GHz is its frequency,
and $L_\odot\sim 3.9\times 10^{33}\,{\rm erg}/$s is the solar bolometric luminosity.
For radio galaxies we simply adopt their typical mass of about $10^{12}\,M_\odot$ to estimate
the Fourier transform of the their density profile.

\subsection{Fourier Transforms}

\begin{figure}
\includegraphics[width=0.6\textwidth]{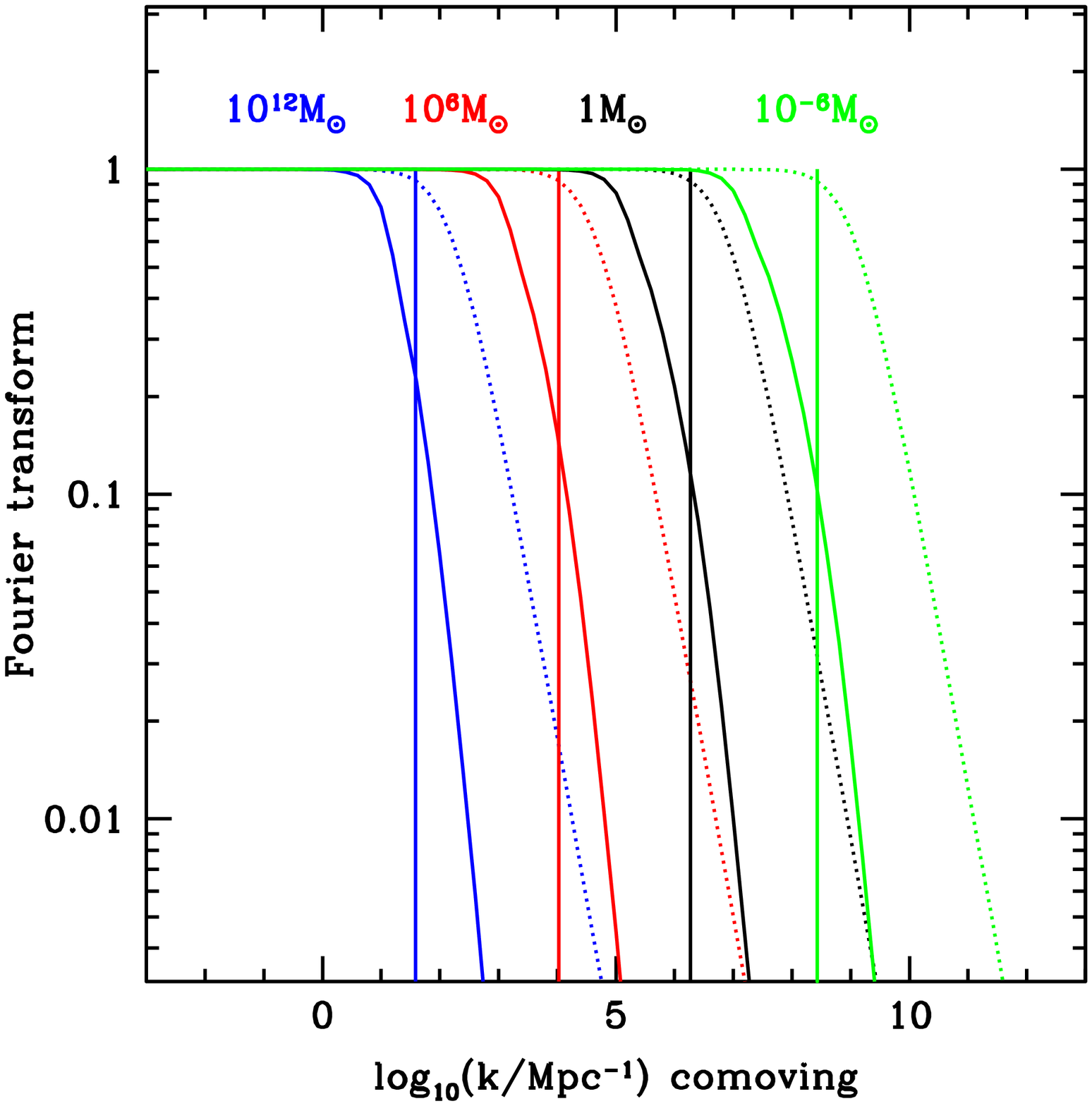} 
\caption{The normalized Fourier transforms $y_1(k,M)$ (solid lines) and $y_2(k,M)$ (dotted lines)
of $\rho_h$ and $\rho_{h^2}$, respectively,
as functions of co-moving wavenumber $k$. The vertical lines denote the scale $k=1/r_s(M)$.}
\label{fig:fourier_rho}
\end{figure}

The Fourier transform of the spherically symmetric NFW profile of mass M can be written as
\begin{equation}
 {\cal F}_{\rho_h}(k,M)=\int_0^{r_v}\rho_h(M,r)\frac{\sin(kr)}{kr}4\pi r^2dr\,,
\end{equation}
and analogously for ${\cal F}_{\rho^2_h}(k,M)$. For the purpose
of plotting these Fourier transforms, see Fig.~\ref{fig:fourier_rho},
it is convenient to renormalize them to unity for $k\to0$ by introducing the new functions
$y_1(k,M)={\cal F}_{\rho_h}(k,M)/M$ and
$y_2(k,M)={\cal F}_{\rho^2_h}(k,M)/\int dV_h \rho_h^2({\bf r})$.
We then have $y_i(0,M)=1$, and $y_i(k>0,M)<1$ for $i=1,2$. For the NFW density profile,
$\int dV_h \rho_h^2({\bf r})= f_cM\rho_{m}\Delta_c(z)$, where $f_c=(c^3/9)\left[1-(1+c)^{-3}\right]/\left[\log(1+c)-c/(1+c)\right]^2$,and $\Delta(z)\sim 200$.
 
For the NFW profile, the mass of the halo within radius $r$ increases $\propto r^2$ for $r\la r_s$,
and then increase logarithmically for $r_s\la r\la r_v$ where $\rho_h(r)\propto r^{-3}$. Therefore,
the dominant contribution to the halo mass comes from $r\la r_s$. Similarly, the annihilation signal
is produced mainly within $r\la r_s$, increasing there $\propto r$, but increases only
$\propto r_s^{-3}-r^{-3}$ for $r_s\la r\la r_v$. 
    
Fig.~\ref{fig:fourier_rho} shows that for $kr_s\ll1$ we have $y_{1,2}\simeq1$, whereas for
$kr_s\gg1$ one has $y_2(k,M)\propto k^{-1}$, and $y_1(k,M)\propto k^{-2}$.

\subsection{Foregrounds}
The radio intensity $I_\nu$ at a given frequency $\nu$ can be expressed in terms of antenna temperature
$T_A(\nu)$ via $I_\nu=2\nu^2k_{\rm B}T_A(\nu)/c_0^2$, where $c_0$ is the speed of light.
Alternatively, $I_\nu$ can be written in terms of the thermodynamic temperature
as the temperature of a blackbody with the given intensity at frequency $\nu$, thus
$I_\nu=2\nu^3/(e^x-1)$, where $x\equiv h\nu/k_{\rm B}T$ with $h$ the Planck constant.
Thus, for power law spectra $I_\nu\propto\nu^\alpha$, $T_A\propto\nu^{\alpha-2}$. In general, the
CMB is expressed in terms of thermodynamic temperature $T$, while Galactic and extragalactic
foregrounds are expressed in term of antenna temperature. Thermodynamic and antenna temperature are
then related by $T=T_A(e^x-1)/x$, and their fluctuations by
$\Delta T = \Delta T_A(e^x-1)^2/(x^2e^x)$. For the CMB,
$x=h\nu/(k_{\rm B}T_{\rm CMB})\simeq\nu/(56.8\,{\rm GHz})$ with the CMB temperature
$T_{\rm CMB} = 2.725\,$K~\cite{Mather:1998gm}. Since we consider frequencies
$\nu\la10\,$GHz in the present paper, $x\ll1$ and thus $T\simeq T_A$ and $\Delta T\simeq\Delta T_A$.

From the definition of $T_A$ we get
\begin{equation}
\fl I_\nu=3.06\times 10^{-25}\left(\frac{\nu}{{\rm GHz}}\right)^2\left(\frac{T_A}{\mu{\rm K}}\right)
\,{\rm erg}\,{\rm cm}^{-2}\,{\rm s}^{-1}\,{\rm Hz}^{-1}\,{\rm sr}^{-1}\,.
\end{equation}
Since for $I_\nu\propto\nu^\alpha$ the power spectrum ${C^{I_\nu}_l}$ of $I_\nu$ at frequency
$\nu$ scales as $\nu^{2\alpha}$, we can express it in terms of the power spectrum $C^{T_A}(\nu^\prime)$
of the antenna temperature $T_A$ at frequency $\nu^\prime$ via
\begin{equation}
\fl\sqrt{C^{I_\nu}_l(\nu)} =3.06\times 10^{-25}\,\left(\frac{\nu}{\nu^\prime}\right)^{\alpha-2} \sqrt{\frac{C^{T_A}_l(\nu^\prime)}{\mu{\rm K}^2}}
\left(\frac{\nu}{{\rm GHz}}\right)^2\,{\rm erg}\,{\rm cm}^{-2}\,{\rm s}^{-1}\,{\rm Hz}^{-1}\,{\rm sr}^{-1}\,.
\end{equation}
Here, $\alpha=-0.9$ for synchrotron emission and $-0.15$ for free-free emissions, respectively~\cite{Tegmark:1999ke}. For the normalization and the dependence on $l$, we adopted the best-fit model from observations at
2.3 GHz~\cite{Giardino:2001jv}. These parametrizations have been used in Fig.~\ref{fig:cl_sum}.

\begin{figure}[th]
\includegraphics[width=0.6\textwidth]{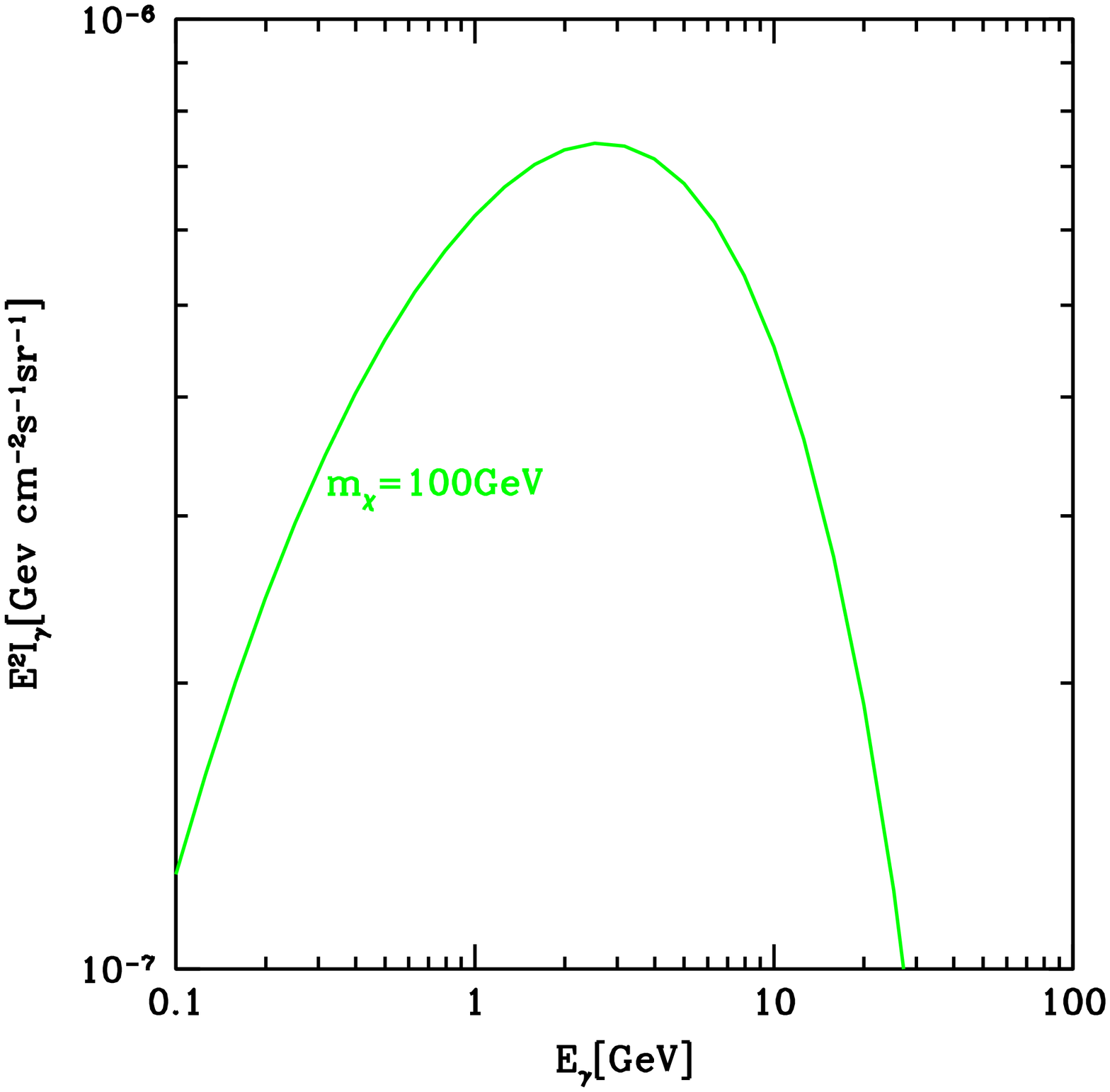} 
\caption{Diffuse energy spectrum of $\gamma-$rays from dark matter annihilation for our fiducial
scenario, with $M_{\rm min}=10^{-6}\,M_\odot$. This is consistent with Fig.~1 in
Ref.~\cite{Ando:2005xg} within about 10\%.}  
\label{fig:I_gamma}
\end{figure} 
\begin{figure}[th]
\includegraphics[width=0.6\textwidth]{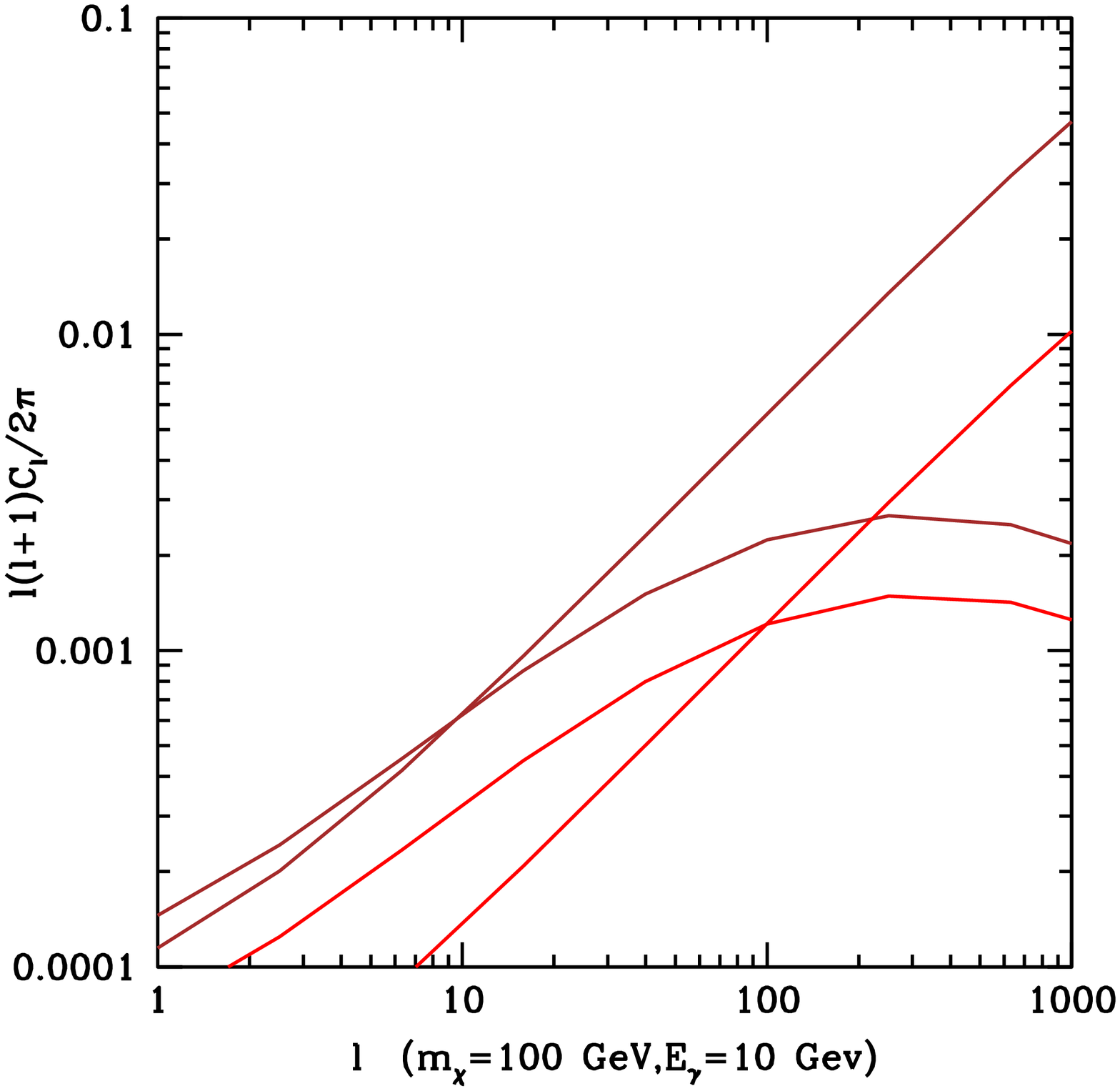} 
\caption{Angular power spectrum of 10 GeV $\gamma-$rays from dark matter annihilation in our fiducial
scenario. Brown lines are for minimal halo mass $M_{\rm min}=10^{6}\,M_\odot$, whereas
red lines are for $M_{\rm min}=10^{-6}\,M_\odot$. The higher and lower
curves at high $l$ denote the one-halo and two-halo terms, respectively. Note that $C_l$ is now
normalized to the total intensity squared, $I_\nu^2$.
This is consistent with Fig.~6 in Ref.~\cite{Ando:2005xg} within a factor $\simeq2$.}
\label{fig:Cl_gamma}
\end{figure}

\subsection{Diffuse Flux and Power Spectrum of $\gamma-$rays}

We can use our approach also to compute diffuse energy spectra and angular power spectra of
$\gamma-$rays from dark matter annihilation. For neutralinos the spectrum of $\gamma-$rays
of energy $E_\gamma$ per annihilation can be parameterized by the simple expression~\cite{Bergstrom:2001jj}
\begin{equation}
  \frac{dN_\gamma}{dE}(E_\gamma)\simeq\frac{0.73}{m_X}
  \frac{e^{-7.776E_\gamma/m_X}}{(E_\gamma/m_X)^{1.5}+0.00014}\,.
\end{equation}
To reasonably match the EGRET data~\cite{Strong:2004ry}, and compare with the results in
Ref.~\cite{Ando:2005xg}, we multiply the predicted average $\gamma-$ray intensity spectrum
by the boost factor $A_b\sim240$ due to substructure within the host halos.
Note that this boost factor is much more extreme than $A_b\simeq10$ assumed
in the present work. We then have
\begin{equation}
  w(E_\gamma,z) = \frac{\langle\sigma v\rangle}{8\pi}
  \left(\frac{\Omega_m}{m_X}\right)^2(1+z)^3\,E_\gamma\frac{dN_\gamma(E_\gamma,z)}{dE_\gamma}
\end{equation}
for the weight function in Eq.~(\ref{eq:limber1}).
For our fiducial dark matter scenario we then obtain
Fig.~\ref{fig:I_gamma} for the solid angle averaged energy spectrum and Fig.~\ref{fig:Cl_gamma}
for the angular power spectrum at $\gamma-$ray energy $E_\gamma=10\,$GeV. Note that if
fainter sources were included, corresponding to smaller minimal halo mass $M_{\rm min}$, the
relative fluctuations normalized to $I_\nu^2$, and especially the one-halo Poisson term, would decrease.

The diffuse energy spectrum is consistent with the results in Ref.~\cite{Ando:2005xg} within
about 10\%. In addition, the angular power spectrum at $\gamma-$ray energy 10 GeV is also consistent
with the results in Ref.~\cite{Ando:2005xg} within a factor two. This serves as an important cross-check of our results with independent calculations.

\end{document}